\documentclass[twocolumn,trackchanges]{aastex62}

\submitjournal{ApJ}

\usepackage{esvect}
\usepackage{savesym}
\usepackage{amsmath}
\savesymbol{iint}
\usepackage{txfonts}
\restoresymbol{TXF}{iint}
\usepackage{graphicx}	% Including figure files
\usepackage{amsmath}	% Advanced maths commands
\usepackage{amssymb}	% Extra maths symbols

\shorttitle{Radio emissions from galaxy groups}
\shortauthors{Paul S., et al.}
\begin{document}

\title{\Large On the prospect of discovering `galaxy groups' through radio observations}

\correspondingauthor{Surajit Paul}
\email{surajit@physics.unipune.ac.in,surajit@associates.iucaa.in}

\author[0000-0002-0786-7307]{Surajit Paul}
\affil{Department of Physics, \\ SP Pune University, Pune 411007, India}

\correspondingauthor{Prateek Gupta}
\email{prateek@physics.unipune.ac.in }

\author[0000-0002-0786-7307]{Prateek Gupta}
\affil{Department of Physics, \\ SP Pune University, Pune 411007, India}

\author{Reju Sam John}
\affiliation{Department of Physics, Pondicherry Engineering College, \\ Puducherry, 605014, India}
\affil{Department of Physics, \\ SP Pune University, Pune 411007, India}

\author{Venkat Punjabi}
\affil{Department of Physics, \\ SP Pune University, Pune 411007, India}

%% Note that the \and command from previous versions of AASTeX is now
%% depreciated in this version as it is no longer necessary. AASTeX 
%% automatically takes care of all commas and "and"s between authors names.

%% AASTeX 6.2 has the new \collaboration and \nocollaboration commands to
%% provide the collaboration status of a group of authors. These commands 
%% can be used either before or after the list of corresponding authors. The
%% argument for \collaboration is the collaboration identifier. Authors are
%% encouraged to surround collaboration identifiers with ()s. The 
%% \nocollaboration command takes no argument and exists to indicate that
%% the nearby authors are not part of surrounding collaborations.

%% Mark off the abstract in the ``abstract'' environment. 
\begin{abstract}

Observed steep mass scaling of radio power from the available high mass galaxy clusters has ruled out the prospect of detection of 'galaxy groups'. On the other hand, the available simulations and observations of thermal emissions show that the groups are merger prone, thus non-virialised, indicating better visibility in the non-thermal radio waves. Detection of radio emissions from them would help us to understand the scale-dependent effectiveness of particle acceleration mechanisms, as well as, being younger and cooler than clusters, groups can be a unique laboratory to test the models of cosmic magnetism. They can also be the potential source of Warm-Hot Intergalactic Medium (WHIM). So, in this study, we have modelled radio emissions from the structures in cosmological hydrodynamic simulations performed using {\sc{ENZO}}. We present a simple model for computing magnetic field using turbulence and for the first time, used the electron energy spectrum from both the diffusive shock acceleration (DSA) and turbulent re-acceleration (TRA) mechanisms to compute radio synchrotron emissions. Computed radio power from a wide range of mass ($\geq 10^{13}$ to $2\times 10^{15} M_{\odot}$) with a sample of more than 200 simulated objects show a new mass scaling of $M_{500} \propto P_{1.4\;GHz}^{2.17\pm 0.08}$ and a strong correlation scale of $L_X \propto P_{1.4\;GHz}^{1.08\pm 0.05}$. Both magnetic field and radio power are shown to have adequately replicated the available observations at high mass, allowing us to extend the results to further smaller masses. We report that groups below $10^{14}\;\rm{M_{\odot}}$ show the existence of 10s of nano-Gauss to a sub-$\mu$G magnetic field and about 10$^{19-23}$ W Hz$^{-1}$ of radio power, much higher than what existing mass scaling predicts. We found that the combined radio power from TRA and DSA electrons can only fit very well to all the observed `radio halos' which significantly improves our understanding about radio halo emission. Finally, we have  implemented this model on a real data set obtained from the Sloan Digital Sky Survey (SDSS). It predicts about 10s to 100s $\mu$Jy/(10$\arcsec$ beam) of radio flux in groups indicating their detectability with existing and aplenty with the future radio telescopes. 

\end{abstract}

%% Keywords should appear after the \end{abstract} command. 
%% See the online documentation for the full list of available subject
%% keywords and the rules for their use.

%% From the front matter, we move on to the body of the paper.
%% Sections are demarcated by \section and \subsection, respectively.
%% Observe the use of the LaTeX \label
%% command after the \subsection to give a symbolic KEY to the
%% subsection for cross-referencing in a \ref command.
%% You can use LaTeX's \ref and \label commands to keep track of
%% cross-references to sections, equations, tables, and figures.
%% That way, if you change the order of any elements, LaTeX will
%% automatically renumber them.
%%
%% We recommend that authors also use the natbib \citep
%% and \citet commands to identify citations.  The citations are
%% tied to the reference list via symbolic KEYs. The KEY corresponds
%% to the KEY in the \bibitem in the reference list below. 

\section{Introduction}\label{intro}
The large scale matter distribution in the universe has an interconnected web like structure that is comprising of galaxies, galaxy groups, large filaments and clusters of galaxies \citep{Jones_2004RvMP,Springel_2006Natur}. In the structural hierarchy, `galaxy groups' (for definition see \citet{Paul_2017MNRAS}) are the intermediate objects between the field galaxies and the rich galaxy clusters \citep{Freeland_2011ApJ}.  Groups, that usually form inside the dark matter (DM) filaments connecting the clusters \citep{lietzen_2012A&A,Tempel_2014A&A_a,Vajgel_2014arXiv}, appear like series of knots on a long string \citep{Tempel_2014A&A_a} and eventually pulled towards the nodes (clusters) \citep{Pimbblet_2011MNRAS,Perez_2009MNRAS,Moss_2006MNRAS}. 
They are thus observed to be stretched along the filaments \citep{Zhang_2013ApJ}, possibly due to their rapid movement towards the cluster. While moving through these DM channels, they would experience tremendous shear force and dynamical friction and get squeezed along the perpendicular direction to the filaments, increasing the rate of collisions among the constituent galaxies \citep{Struck_2011gaco.book,Diaferio_1993AJ,O'Sullivan_2015A&A}. 

Groups are mostly found to be non-virialized because, they are very much unstable to mergers as shown in N-body simulations \cite{Carnevali_1981ApJ,Diaferio_1993AJ}), in recent work by \cite{Diaz_2010MNRAS}, and in hydrodynamical simulations by \cite{Paul_2017MNRAS}. Moreover, being inside the shallower gravitational potential of filaments, they are expected to be more strongly affected by processes such as mergers, feedback from super-massive black holes (SMBH), and galactic winds etc. \citep{Lovisari_2015A&A}. Fractional cosmic ray content that mainly depends on Mach number of shocks in the inter-galactic medium (IGM) is also reported to be larger in some galaxy groups compared to the clusters \citep{Jubelgas_2008A&A,John_2018arXiv}. In a recent work by \citet{Paul_2017MNRAS}, it has been shown that the CR luminosity follows a different evolutionary path than that of the clusters and fluctuations of CR luminosity in low mass systems observed to be higher in their merging state \citep{John_2018arXiv}. The frequent merger would introduce turbulence and may allow groups to re-accelerate particles and convert more thermal energy to non-thermal energy. This indicates that non-thermal scaling laws derived from the cluster observations may not be obeyed by the groups.  In fact,  such deviation has also been reported by \citet{Paul_2017MNRAS}. So, the study of galaxy groups can provide vital  information of an intermediate environment between the field galaxies and the clusters \citep{Berrier_2009ApJ}, especially in non-thermal emissions. 

In recent years, the importance of low mass objects in understanding large scale structure (LSS) formation and its energy evolution has been pointed out by various researchers  \citep{Paul_2017MNRAS,Bharadwaj_2015A&A,Vajgel_2014arXiv,Pratt_2009A&A}, But, their non-thermal properties, especially the radio emissions remained almost unexplored. Since, non-thermal radiations depend more on the transient activities such as mergers, shocks, and turbulence etc., it can reveal much more information about the dynamics of LSS (for review \cite{Dolag_2008SSRv,Bruggen_2012SSRv}) over any other form of energies. But, groups are not sufficiently observed yet in non-thermal emissions and no general properties are studied in this regard. Cluster mass-scaling with existing radio observations \citep{Cassano_2013ApJ,Yuan_2015ApJ} indicate a steep slope of about 4, removing the possibility of detection of galaxy groups (mass $< 10^{14} M{\odot}$) with currently available telescope facilities. But, as the data available or used in these studies are incomplete (only for massive clusters $\gtrsim 5 \times 10^{14} M{\odot}$), it may not be very wise to extrapolate this to a wide range of masses. \citet{Paul_2017MNRAS} shows, turbulence and cosmic rays from group significantly deviates from cluster scaling and becomes flatter, indicating better visibility of groups through their non-thermal properties. This motivated us to model the radio emissions from smaller objects through simulations.

In this study, we have performed cosmological hydrodynamic plus N-body simulations using {\sc{Enzo}} code \citep{Bryan_2014ApJS} with a specific aim to understand the non-thermal radio emissions from objects at different scales ranging more than two orders of magnitude in mass ($10^{13}M_{\sun}-10^{15} M_{\sun}$). We have theoretically modelled LSS taking into account of radiative cooling, heating due to star motions and supernova and star formation feedback physics. AGN feedback may have a non-negligible effect in the core of very low mass ($\sim 10^{13} M{\odot}$) systems \citep{McCarthy_2010MNRAS}, but this is out of scope of this work and thus becomes a topic for future study (see details of our used model in \citet{Paul_2017MNRAS}). Using the above model, we have created a mock sample of more than 200 objects for our study. We have the smaller clusters and galaxy groups in plenty that are not yet explored well in radio waves, leaving a vacuum of information in between field galaxies and the clusters. The wide range of mass in our study helped us to compare our simulated objects with the available observations and to constrain our model which in turn, made us able to extrapolate our results to the smaller objects more accurately. Further, our model has been applied to compute possible radio emissions from real objects in the SDSS galaxy group list \citep{Tempel_2014A&A}. Finally, we have computed possible observable fraction of these SDSS group sample for the available and upcoming highly sensitive radio telescopes such as uGMRT, SKA and so on. 

This article has five sections. After giving an introduction in Section~\ref{intro}, we have described our simulation details and mentioned about computed and observed data samples in Section~\ref{Selected-data-sets}. Our model of finding turbulence, computing magnetic fields and radio emissions from different particle acceleration mechanisms are described in Section~\ref{radio-model}. We have discussed our simulation results and implementation of our model on the observed SDSS sources with the possibility of observations in Section~\ref{res}. Finally, results have been summarised and concluded in the Section~\ref{sum}.

\section{Data selection details}\label{Selected-data-sets}

\subsection{Simulation details and the data}\label{simu}

To understand and make a possible theoretical estimation of non-thermal radio emission from groups, we have created our sample of objects by performing simulations with the adaptive mesh refinement (AMR), grid-based hybrid (N-body plus hydro-dynamical) code Enzo v.~2.2 \citep{Bryan_2014ApJS,O'Shea_2004astro.ph}. A (128 Mpc)$^3$ volume has been simulated with the introduction of 2 nested child grid and further 4 levels of AMR at the central (32 Mpc)$^3$ volume, a resolution of about 30 kpc has been achieved at the highest resolved level. As cosmological parameters, we have taken a flat $\Lambda$CDM background cosmology with $\Omega_\Lambda$ = 0.7257, $\Omega_m$ now = 0.2743, $\Omega_b$ = 0.0458,  h = 0.702 and primordial power spectrum normalization $\sigma_8$ = 0.812 \citep{Komatsu_2009APJS}. 

Since, shocks and turbulence are the two most important parameters that are required to compute radio emissions (Further details in Section~\ref{radio-model}), we paid more attention to these parameters in our simulations. The shocks have been computed in our simulations using un-split velocity jump method of \citet{Skillman_2008ApJ} with a temperature floor of $\rm{T}^4$K. This method is found to produce better results in AMR simulations \citep{Vazza_2011MNRAS}. Since shocks are very much important mainly for DSA computation, in our adaptive mesh refinement strategy, the refinement criteria based on shocks along with the over-density (both in DM and baryon) have been used. A detailed account of the AMR criteria used here can be found in \cite{John_2018arXiv}. Turbulence, on the other hand, is a derivative of kinetic energy due to dynamics of the systems which is in turn controlled by the distribution of different forms of energies in the inter cluster and inter galactic medium. So, to obtain a more realistic energy distribution in the LSS, a formulation has been used that includes the effect of radiative cooling due to X-ray, UV \& optical emissions and heating due to stellar motions and Supernova \citep{Sarazin_1987ApJ}. We have also used the star formation and feedback scheme of \citet{Cen_1992ApJL} with a feedback of 0.25 solar in our simulations. In short, we call this model with additional physics as `coolSF' runs (for details, see \citet{Paul_2017MNRAS}).

With the above-said model, we have simulated 10 realizations of (128 Mpc)$^{3}$ volume. We have used the 30 kpc resolution simulations as the reference set (`REFRES' hereafter). Each simulation has been started at redshift z=60 and ran till z=0. We have taken several snapshots at different redshifts, mostly at the low redshift (below z=1). Further, computations of different physical properties for this study have been performed as the post processing of these simulated snapshots by using the yt-tools \citep{Turk_2011ApJS}. 

\subsubsection{Simulated data set}\label{sim-data}

\begin{figure}
\includegraphics[width=0.52\textwidth]{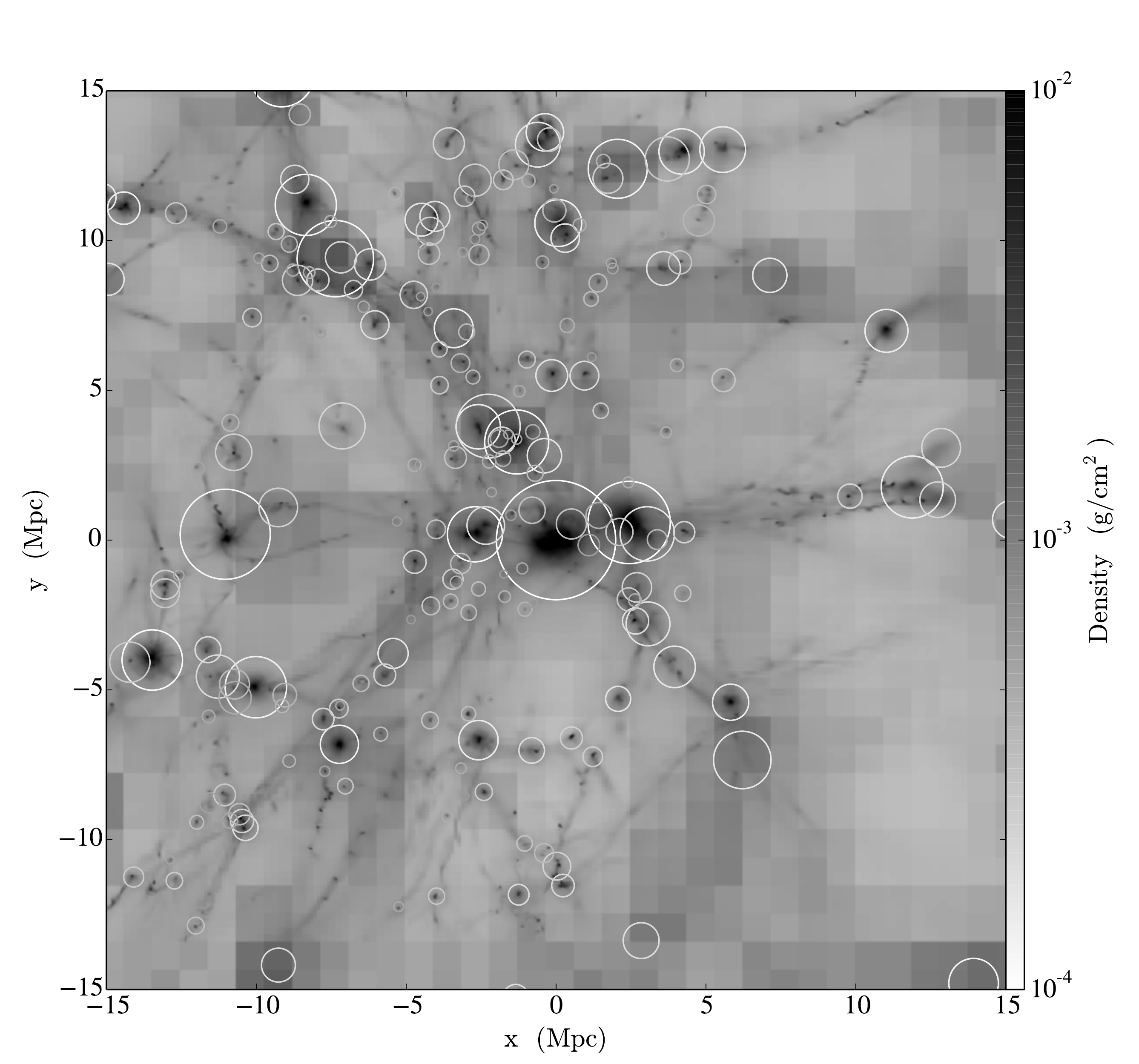}
\caption{Baryon density plotted for (30 Mpc)$^2$ simulated area. White circles indicate the structures and circle radius is same as their virial radius i.e. $r_{200}$.}\label{gr-filaments}
\end{figure}

From one of our simulations, a slice of $(30 Mpc)^2$ area has been plotted in Figure~\ref{gr-filaments}. It shows a filamentary structure around few massive clusters shown as bigger white circles. Smaller circles representing groups and are observed to be in abundant numbers and placed mostly along the filaments in accordance with the observations as discussed in section~\ref{intro}. 

We have identified groups and clusters from our simulations using the HOP algorithm \citep{Eisenstein_1998ApJ} implemented in yt \citep{Turk_2011ApJS}. The initial sample was selected depending on their $M_{500}$ mass. Where $M_{500}$ is the mass within the radial distance where over-density of the objects become 500 times the critical density of the universe at that redshift. In this scale, usually `virial radius' refers to the radius at over density of 200. Each of our (128 Mpc)$^{3}$ volume simulations were focused on a central cluster, but, a large number of other objects are available within the (32 Mpc)$^3$ central high resolution volume where AMR has been allowed. So, adding to the main 10 clusters, we have a few hundreds of objects with the same resolution when we considered the data from different sets of simulations and from different snapshot outputs. To make sure that the DM particles and baryon gas that forms these objects of interest are mostly coming from the well refined Lagrangian region, we have taken the objects only within central (20 $Mpc)^3$ volume. Finally, we have chosen over 200 groups and clusters with masses above $10^{13}\; M_{\odot}$ and it spans till $2\times10^{15}\; M_{\odot}$. Our mass resolution at the smallest child grid is $<$ $10^{9}\; M_{\odot}$ providing enough mass resolution for the groups with at least 10$^{4}$ particles. Also, with $\sim$ 30 kpc spatial resolution, systems above $10^{13} \; M_{\odot}$ at $r_{500}$ that have virial radii above 500 kpc provide adequate spatial resolution with at least 1000s of cells.

\subsection{Observed data set from SDSS group catalogue}\label{obs-data}

Along with the simulated data, we have chosen a set of observed data from Sloan Digital Sky Survey (SDSS). Model devised through simulations has been implemented to this data set to make an estimation of expected radio flux from real objects. Our primary observed data source is the `galaxy group' catalogue of \cite{Tempel_2014A&A} prepared using the SDSS, Data Release-10. On the tabulated galaxy data, authors applied Friends of Friends (FoF) algorithm, and identified a total of  82,458 galaxy groups by defining groups as an object having at least 2 galaxies and choosing suitable linking length (see \cite{Tempel_2014A&A} for details). We know that a group of galaxies is a gravitationally bound system with a mass of approximately $\rm{10^{13} M_{\odot}} $ and with a radius of just less than a Mpc and temperature $\lesssim$ 1 keV \citep{Paul_2017MNRAS}. So, to ensure the group properties, we have chosen only those objects that are having at least 10 numbers of galaxy candidates and having total mass greater than $10^{13} M_{\odot}$ to remove the possibility that groups properties get dominated by merely a few galaxies. We have only put constraint on minimum mass and kept all the higher mass objects (clusters) in the list to compare them with the available observations as well as simulations. So, in our sample, we got 2300 objects with masses in the range of $10^{13} - 2\times 10^{15} M_{\odot} $, the same range as the simulated sample set that we are working with. 

\section{Modelling radio synchrotron emission from groups and clusters}\label{radio-model}

Radio synchrotron emission strongly depends on the energy distribution of available charged particles in astrophysical plasma as well as the magnetic field in the medium \citep{Feretti_2008LNP,Rephaeli_2015ApJ,Kale_2016JApA}. The power spectrum of radio emission is determined by the energy spectrum of the radio emitting electrons which is determined by the particle acceleration mechanisms active in these objects. In large scale astrophysical systems, the known major particle acceleration mechanisms are Diffusive Shock Acceleration (DSA) i.e. Fermi I \citep{Drury_1983RPPh} and Turbulent Re-acceleration (TRA) i.e. Fermi II \citep{Brunetti_2007MNRAS} mechanism). Shocks and turbulence in the IGM are also very important as they determine the strength of magnetic fields either by amplification by compression \citep{Iapichino_2012MNRAS} or by the turbulent dynamo \citep{Subramanian_2006MNRAS}. 

\subsection{Shocks and turbulence in groups and clusters}
\subsubsection{Shocks in groups and clusters}\label{shocks}

Structure mergers can release binding energy of about $10^{61}$ ergs (groups of mass of $10^{13} M_{\odot}$) to  $10^{65}$ ergs (clusters of mass of $10^{15} M_{\odot}$). It has been proposed that the energy released during these mergers create huge pressure in the core of the objects and the medium eventually starts expanding supersonically like a blast wave, inducing strong shocks in the baryonic gas \citep{Ha_2017arXiv,Paul_2012JoPConS,Sarazin_2002ASSL} and travels radially as spheroidal wave-front towards the virial radius and moves beyond \citep{Weeren_2011MNRAS,Machado_2013MNRAS,Ha_2017arXiv,Paul_2011ApJ,Iapichino_2017MNRAS}.  
Shocks then dissipate energy through heating and turbulence stirring in the medium \citep{Sarazin_1987ApJ,Dolag_2005MNRAS,Paul_2011ApJ}.  Also, due to shock acceleration, charged particles in the thermal plasma gets accelerated to high energies.
Time-scale of such dissipation for a single merger is found to be about 1-2 Gyr \citep{Paul_2011ApJ,Roettiger_1999ApJ} indicating an energy dissipation rate of the order of 10$^{47}$   $erg\; s^{-1}$ \citep{John_2018arXiv}. If a few percent of this energy goes to radio emissions, it amounts to at least about 10$^{45}$   $erg\; s^{-1}$.

\begin{figure}
\includegraphics[width=1.1\columnwidth]{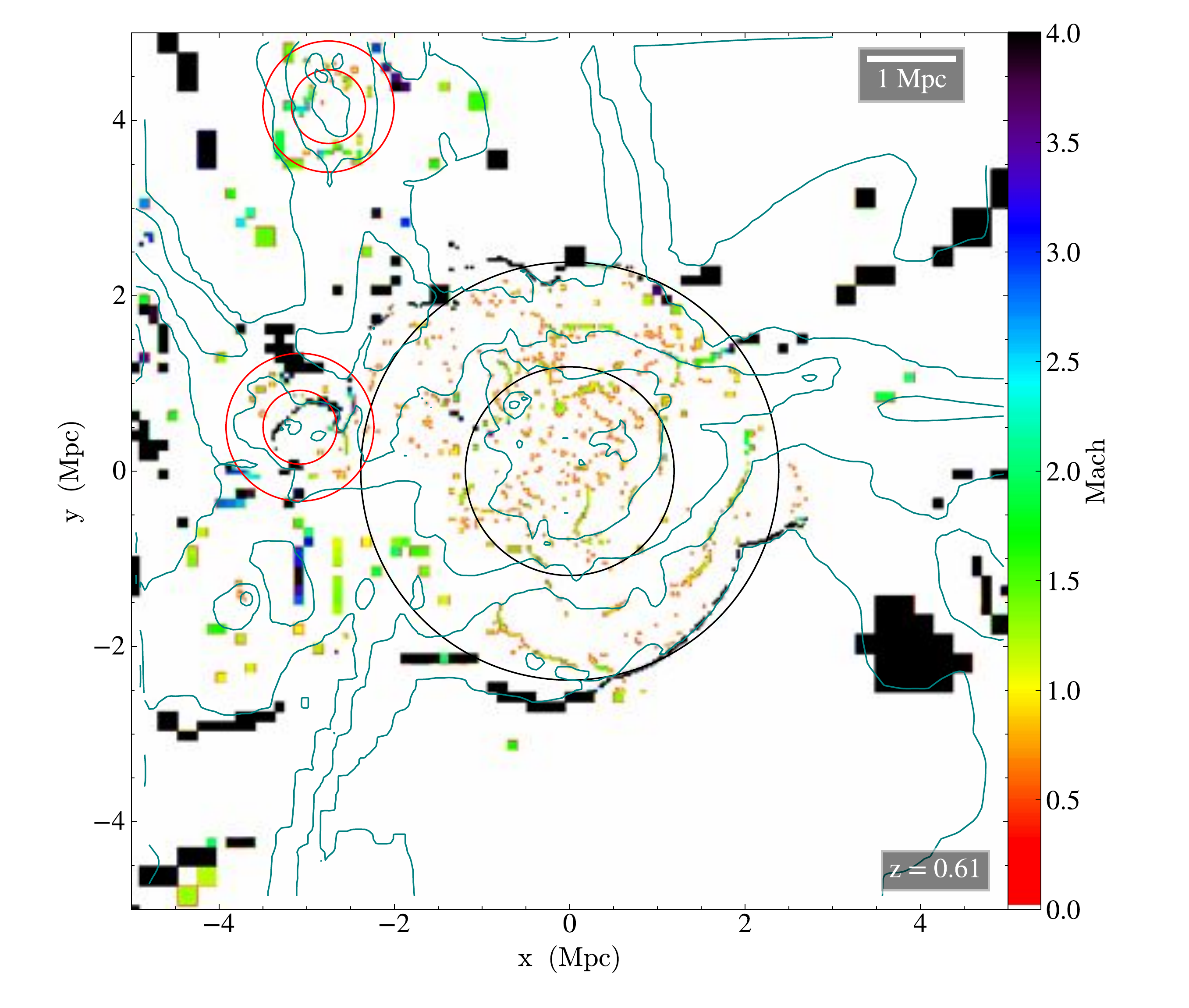}
\hspace{-.5cm}
\caption{Mach number of shocks has been plotted as colour map. Black circles are showing the $r_{1000}$ (inner) and $r_{200}$ (outer) of a big cluster. Same for the groups are shown as red circles.}\label{mach-gr-cl}
\end{figure}

In our simulated objects (Fig.~\ref{mach-gr-cl} is shown as a representative map), in general it is found that in the clusters (indicated by large black circles), shock Mach numbers varying in the range $\mathcal{M}=1-2$ in the core region i.e inner circle at $r_{1000}$. It goes to almost $\mathcal{M}=4$ in regions beyond the core but inside virial radius i.e. outer circle at $r_{200}$, it rarely reaches as high as $\mathcal{M}=10$. These are usually known as the merger or the internal shocks \citep{Miniati_2000ApJ,Skillman_2008ApJ}. Outside the virial radius (i.e.  $r_{200}$), it goes beyond $\mathcal{M}=10$ due to continuous accretion of matter known as accretion or external shocks. Whereas, in the groups, depicted as smaller concentric red circles, we hardly see any shocks inside $r_{1000}$ i.e. inner circle.

\subsubsection{Turbulence in groups and clusters}\label{turb-in-gg-gc}

\begin{figure*}
\includegraphics[width=0.52\textwidth]{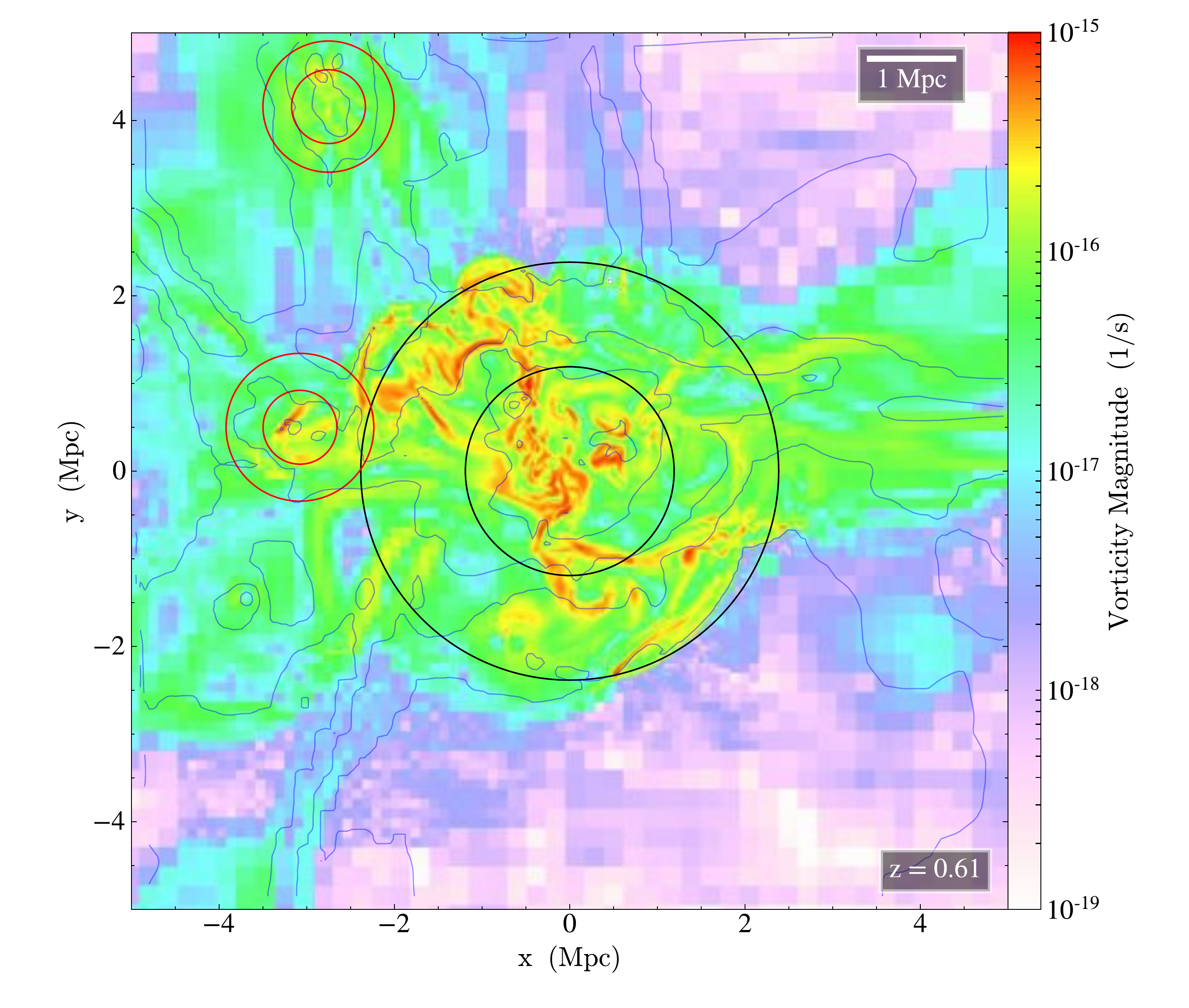}
\includegraphics[width=0.52\textwidth]{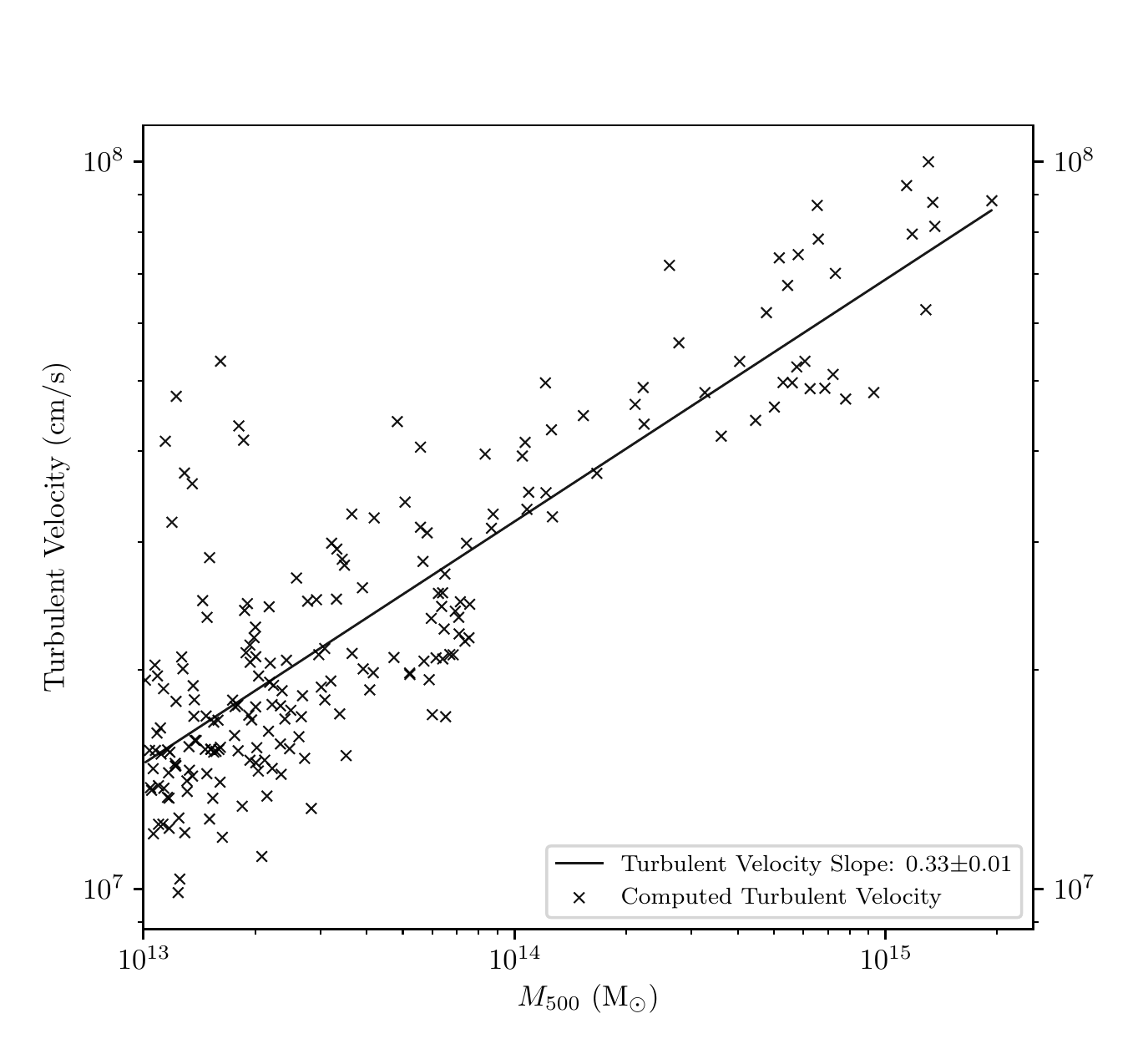}
\caption{{\bf Panel 1:} 20 Mpc$^2$ simulated area with vorticity in colour. Circles are representing clusters and groups similar to Figure~\ref{shocks}. {\bf Panel 2:} Velocity dispersion of baryons with the mass ($M_{500}$) of the objects from our simulated data set.}\label{vorti-groups}
\end{figure*}

Galaxy groups while flowing along the filaments, shown to inject high level of vorticity ($\omega = \nabla \times v$) (see Fig.~\ref{vorti-groups}, Panel 1). It can be noticed that while the clusters are having the highest level of vorticity magnitude about $5 \times 10^{-16}\:s^{-1}$, groups are not far behind. The yellow-green patches along the filaments in Fig~\ref{vorti-groups} Panel 1, are the groups (marked with smaller red circles) and are having vorticity varying in the range of few times 10$^{-17}$ to even $3\times$ 10$^{-16}\:s^{-1}$, values not very far compared to the clusters. 

Turbulent energy in cluster medium is dominated by ($\sim$ 90-95\%) the solenoidal mode which comes from the rotational motions in the IGM \citep{Miniati_2015Natur,Vazza_2017MNRAS}. So, a fluid with non-zero vorticity $\bf{\rm{\vv{\omega}} = \nabla\times \rm{\vv{v}}}$ will have non zero values of turbulent energy. The direction of $\bf{\rm{\vv{\omega}}}$ being random, vectorial average tends to be zero. Therefore the magnitude of vorticity is an important factor for calculation of turbulent energy. Enstrophy, given by $\xi = \int_S \omega^2 {\bf \rm{dS}}$. $\xi$, which is an integral of square of vorticity over a surface $S$, is a proxy to the dissipation of kinetic energy to turbulent every \citep{Zhu_1997ApScR,Vazza_2017MNRAS,Iapichino_2010AIPC}. So, it can be related to the turbulent flow by the Kinetic energy content per unit mass $\epsilon_{turb} \approx 1/2 (v_{rms}^2)$, where $v_{rms}$ i.e. velocity dispersion of the medium. This tells us that the quantification of local velocity dispersion i.e. root mean square velocity ($\rm{v}_{rms}$) would suffice to estimate turbulent energy. 

In this study, $\rm{v}_{rms}$ has been computed by filtering out the bulk motions of baryons in the core ($r_{1000}$) of the groups from our simulated objects (also see \citet{Paul_2015fers.confE}). Finally, this allowed us to compute the turbulent energy of the dominant solenoidal flow mode. A volume weighted spherically averaged quantity within $r_{500}$ is computed to get the average turbulent energy from each object. With this simple method of computing the local turbulent velocity, we got average values of about 100 to 300 km~s$^{-1}$ in our simulated groups (up to the mass $10^{14}M_{\odot}$) that corroborates the observed values well \citep{Rasmussen_2006MNRAS,Wilman_2005MNRAS,Tully_1987ApJ}. But, a noticeable fraction of the groups are found to be very active with even much higher values of turbulence at 300-600 km~s$^{-1}$, almost at the level of clusters (mass beyond $10^{14}M_{\odot}$) which shows values varying in the range 300-1000 km~s$^{-1}$ (Fig.~\ref{vorti-groups}, Panel 2). This is very much in accordance with our expectations (see Section~\ref{intro}).

\subsection{Computing magnetic field}\label{mag-field-compt}

Turbulent energy increases the particle energy stochastically \citep{Brunetti_2007MNRAS,Donnert_2014MNRAS} or gets converted to magnetic energy through turbulent dynamos \citep{Subramanian_2006MNRAS,Rincon_2016PNAS}. It has been reported that magnetic field of $\mu$G that we usually observe today \citep{Govoni_2004IJMPD} in the inter cluster medium (ICM), can be achieved through this turbulent dynamos \citep{Xu_2009ApJ,Ryu_2008Sci}. This mechanism can boost the level of IGM seed magnetisation of about $10^{-21}$ Gauss at re-ionisation era to the present value of $\mu \rm{G}$. Such kind of amplification of magnetic field certainly needs a very high degree of turbulence in the medium i.e. a fully developed or Kolmogorov type, $E(k) \propto k^{-5/3}$ \citep{Egan_2016arXiv}. A quasi-equipartition is reached between magnetic energy density $\frac{B^2}{8\pi}$ and the kinetic energy density $\rho \epsilon_{\rm{turb}}$ in such a fully turbulent medium \citep{Miniati_2015Natur}. So, the saturated magnetic field can easily be obtained from the available hydrodynamic parameters in our simulations by the below given relation

\begin{equation}
\frac{B_{\rm{sat}}^2}{8\pi} \propto \rho \epsilon_{\rm{turb}} i.e. B_{\rm{sat}} = \sqrt{{C_E}.8\pi. \rho \epsilon_{\rm{turb}}}
\end{equation}
 \citep{Subramanian_1998MNRAS,Iapichino_2012MNRAS} where, $\epsilon_{\rm{turb}}$ is the local turbulent energy which is a fraction of the kinetic energy of the medium. The constant of proportionality $C_E$ can be at the max 0.05 \citep{Miniati_2015Natur}.

\subsection{Computing synchrotron radio emissions}\label{synch-rad-compt}

\subsubsection{Radio emission due to DSA}\label{DSA-compt}

The ICM and IGM are shocked due to mergers or accretion of mass clumps during  large scale structure formation (see section~\ref{shocks}). Shocks in the clusters, thermalizes the ICM to as high as $10^{7-8}K $.  Calculation using Saha-ionisation equation \citep{Saha_1920PhilMag} shows this medium to be fully ionised. These highly energetic thermal electrons then pumped into the shocks. Injected charge particles gain energy and get accelerated by crossing these shocks multiple times (Fermi-I mechanism or DSA) and produce a power law energy spectrum with relativistic energy given by $n(E) dE \propto E^{-\delta} dE$ \citep{Drury_1983RPPh,Baring_2009AIPC}, where $\delta$ is the spectral index which is related to the density compression ratio, C, as $ \delta = (C + 2)/(C - 1)$ \citep{Hoeft_2007MNRAS} where density compression ratio can be obtained by Mach number ($\mathcal{M}$) of the shocks as $C= 4\mathcal{M}^2/(\mathcal{M}^2+3/2)$ \citep{Drury_1983RPPh} with polytropic index $\gamma=5/3$. So, final spectral index is a strong function of shock Mach number. These electrons then gyrate in the magnetic field compressed by these shocks or amplified by the turbulence and eventually radiates synchrotron radio emission from the shock surfaces. In this respect a well used relation for radio power due to DSA electron has been derived by \citet{Hoeft_2007MNRAS}

\begin{eqnarray}
\frac{dP(\nu_{obs})}{d \nu} &=& 6.4 \times 10^{34} \frac{erg}{sHz} \left(\frac{A}{Mpc^{2}}\right) \left(\frac{n_e}{10^{-4}cm^{-3}} \right) \left(\frac{T}{7keV} \right)^{\frac{3}{2}} \nonumber \\
                             & & \times \left(\frac{\xi_{e}}{0.05} \right) {\left(\frac{\nu_{obs}}{1.4\;GHz}\right)}^{-\frac{\delta}{2}} \frac{\left(\frac{B}{\mu G} \right)^{1 + \frac{\delta}{2}}}{\left(\frac{B_{CMB}}{\mu G} \right)^{2} + \left(\frac{B}{\mu G} \right)^{2}} \Psi(\mathcal{M})
\end{eqnarray}

where $A$ is the area of the shock wave, $B$ is magnetic field, $B_{CMB}$ is the magnetic field corresponding to the energy density of Cosmic Mircowave Backgroud Radiation (CMBR), $\xi_{e}$ is the electron acceleration efficiency, $\nu_{obs}$ is the observed frequency, $n_e$ is the post-shock electron density, and $T$ is post-shock temperature. $\Psi(\mathcal{M})$ is a dimensionless function of shock Mach number. For strong shocks $(\mathcal{M} \geqslant 5)$, it saturates to $\Psi \sim 1$, while for weak shocks $(\mathcal{M} \leqslant 3)$, $\Psi$ function dies out rapidly \citep{Hoeft_2007MNRAS}. 

In our simulations, we have obtained the post-shock temperature and computed the Mach number of the shocked cells using the unsplit velocity jump as described by \citet{Skillman_2008ApJ}. Post-shock electron density is being computed on each cell as post process derived quantities. Shocked surface area has been approximated here as the sum of the area of one surface of each of the cells tagged as shocked (See \citet{Vazza_2017MNRAS}). Computation of radio power through DSA is being done on each cell in shocked regions with having Mach number greater than $\mathcal{M}$=1.5 as the radio emission shown to take place only beyond Mach $\mathcal{M}$=1.5. It has also been noticed that power emitted by shocks rapidly falls beyond Mach $\mathcal{M}$=4.5 \citep{Hong_2014ApJ}. The Mach number range of $\mathcal{M}$=1.5-4.5 is also found to be the most effective cosmic ray producer \citep{John_2018arXiv}, indicating more availability of synchrotron electrons.

\subsubsection{Radio emission due to TRA}\label{TRA-compt}
Turbulence generated in forming structures (mainly during mergers) as discussed in section~\ref{turb-in-gg-gc} can re-accelerate ambient energetic charge particles stochastically by the process known as the Fermi II or the Turbulent Re-Acceleration (TRA) \citep{Brunetti_2001A&A,Brunetti_2007MNRAS}.
Induced turbulence in the IGM generates Alfven waves through Lighthill radiation \citep{Lighthill_1952RSPSA}. These Alfven waves accelerate ambient high energy electrons to GeV energies. The highest energy electrons those could resonate with the turbulence is given by 
\citep{Fang_2016JCAP} as;

\begin{eqnarray}
E_{\rm max} &=& 53 GeV \quad{\left(\frac{B}{\mu G}\right)}^{\frac{4}{3}} {\left(\frac{T}{2 keV}\right)}^{\frac{-1}{6}} {\left(\frac{l_0}{300 kpc}\right)}^{\frac{2}{3}} \nonumber \\
            & & \times \quad {\left(\frac{v_t}{300 kms^{-1}}\right)}^{\frac{-4}{3}} {\left(\frac{n_e}{10^{-3} cm^{-3}}\right)}^{\frac{-1}{6}} 
\end{eqnarray}

$v_t$ is the turbulent velocity which we have computed as discussed in section~\ref{turb-in-gg-gc} and $l_0$ is the largest eddy size of the system (or the largest turbulent scale in the system) which can be defined as $l_0$ = $V_T \tau_0$, where $V_T$ is the velocity difference of largest eddy scale and $\tau_0$ is the Hubble time \citep{Stein_1974A&A}. As a first order approximation, we have used the most probable velocity within the individual groups and clusters as the value for $V_T$. Usually, turbulence life time in the groups and clusters are not the Hubble time, rather it is the time of sustenance of the turbulence in the system after has been induced by the mergers. We have thus taken $\tau_0$ to be about 2 Gyr as reported by \citep{Paul_2011ApJ}. This gives us well comparable largest eddy scales as estimated by \citet{Subramanian_2006MNRAS,Goldman_1998IAUS}.

In TRA, the particle energy spectrum takes up the form 
\begin{equation}
\left(\frac{dn_e}{dE_e} \right) = \frac{3P_A\,c}{4 S(E_{\rm max})^{1/2}}\,E_e^{-\delta}
\end{equation}

 \citep{Fang_2016JCAP}, where, $P_A$ is the part of the total turbulent power going into the Alfven waves and is given by 
\begin{eqnarray}
P_A &=& \quad 4.2 \times 10^{-32} erg^{} cm^{-3} s^{-1} {\left(\frac{v_t}{300kms^{-1}}\right)}^{\frac{23}{6}} {\left(\frac{B}{\mu G}\right)}^{\frac{-4}{3}} \nonumber \\
    & & \times \quad {\left(\frac{T}{2 keV}\right)}^{\frac{-1}{12}} {\left(\frac{n_e}{10^{-3} cm^{-3}}\right)}^{\frac{5}{3}} {\left(\frac{l_0}{300 kpc}\right)}^{\frac{-7}{6}}
\end{eqnarray}

and $\delta = \frac{5}{2}$ with an assumption of Kolmogorov type i.e. fully developed turbulence, and
 
\begin{eqnarray}
S &=& \frac{4(B^{2} + B_{CMB}^{2})e^{4}}{9m_e^{4}c^{6}}
\end{eqnarray}

where $Sp^{2}c$ corresponds to the synchrotron and inverse Compton emission power of an electron \citep{Fang_2016JCAP}. Using the above particles energy spectrum, the radio synchrotron power through turbulent re-acceleration mechanism computed as 

\begin{eqnarray}
 \frac{d^{2}P(\nu_{obs})}{dVd\nu} &=& \frac{\sqrt{3}e^3B}{8m_e c^2}\,\int_{E_{\rm min} }^{E_{\rm max}}dE_e\,F\left(\frac{\nu_{obs}}{\nu_c}\right)\,\left(\frac{dn_e}{dE_e} \right)_{\rm inj}
\end{eqnarray}

where, $\nu_c$ is the critical frequency of synchrotron emission, 

\begin{equation}
\nu_c =  \frac{3\gamma^2 eB}{4\pi m_e} = 0.016\left(\frac{B}{1\mu G}\right) \left( \frac{E_e}{1 GeV}\right)^2~\rm GHz 
\end{equation}

 and 
\begin{equation}
 F(x)=x\int_x^\infty K_{5/3}(x')dx'
\end{equation} 

 is the synchrotron emission function, which peaks at $x = 0.29$, $K_{5/3}$ is the modified Bessel function of order 5/3 \citep{Longair_2011}.

\section{Results and discussions}\label{res}

\subsection{Computed magnetic fields}\label{mag-field-res}

\begin{figure}
\includegraphics[width=0.52\textwidth]{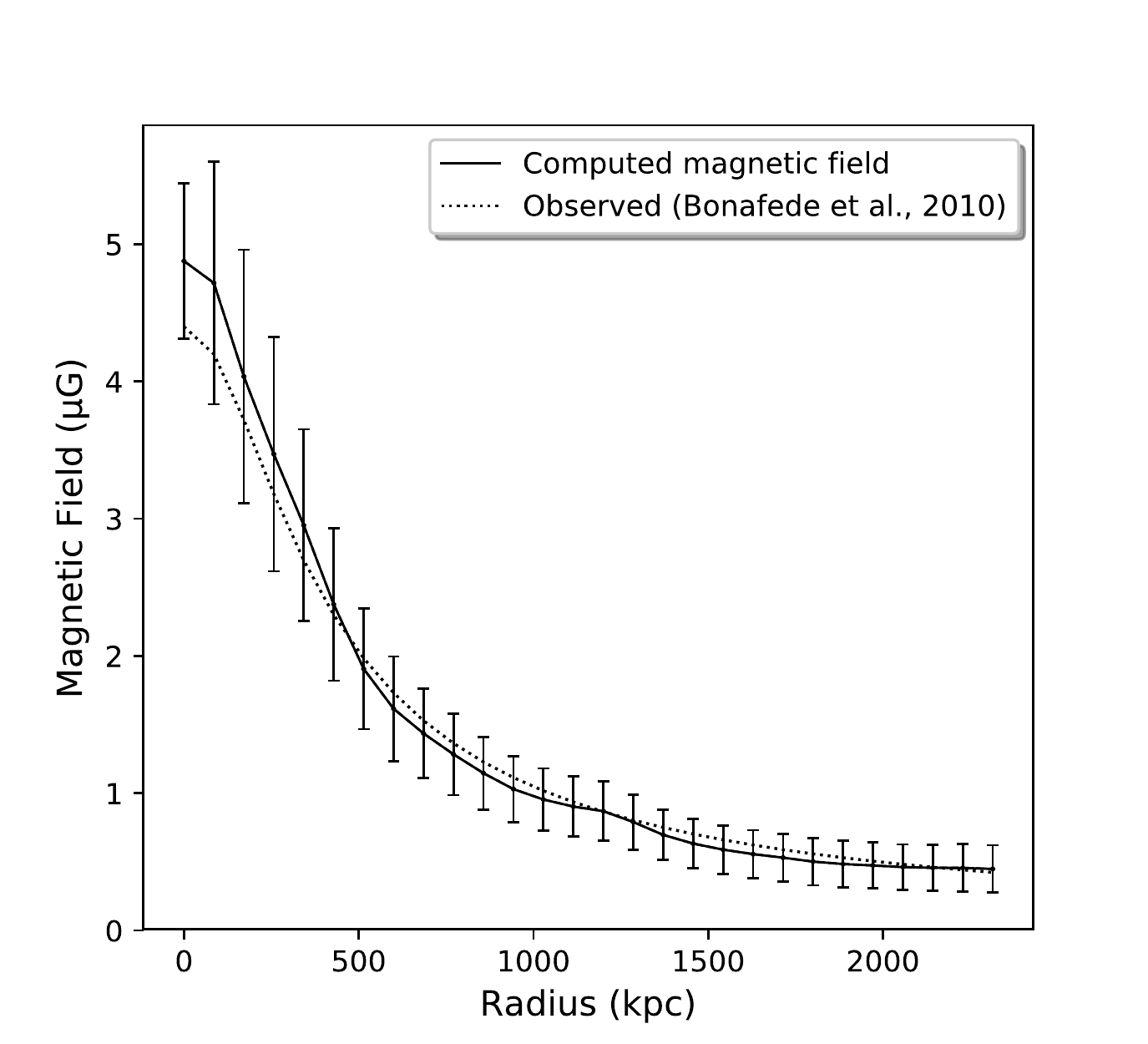}
\includegraphics[width=0.52\textwidth]{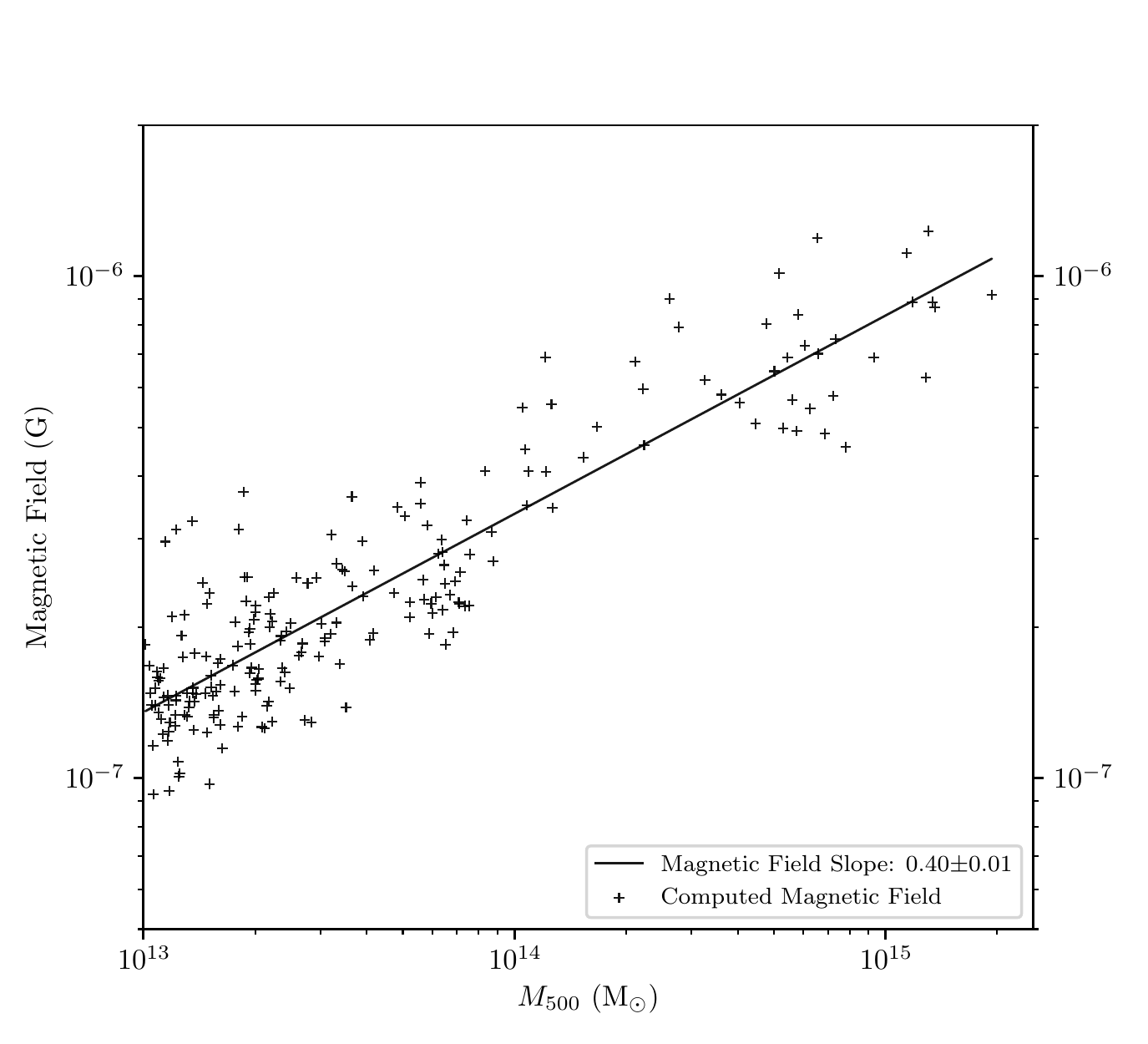}
\caption{{\bf Panel 1:} Computed radial magnetic field over-plotted on observed Coma cluster magnetic field. {\bf Panel 2:} Computed magnetic field for the selected samples plotted against the mass ($M_{500}$).}\label{mag-field}
\end{figure}

We have implemented the model described in section~\ref{mag-field-compt} on a Coma-like cluster with similar mass, radius (about $M_{200}$ about $10^{15}M_{\odot}$, $r_{200}$ about 3 Mpc; within the error bars of \citet{Kubo_2007ApJ,Brilenkov_2015arXiv}) and dynamical state (relatively relaxed \citep{Kent_1982AJ}, with presence of a few sub-clumps) in our simulation. Our modelled magnetic field is in good agreement with the radial profile of Coma cluster plotted till $r_{500}$ using Faraday rotation measurements \citep{Bonafede_2010A&A} (see Fig.~\ref{mag-field}, Panel 1). Further, using the same model, we have computed the magnetic fields for all the objects in our sample used for this study and found average magnetic field $\left<B\right>$ of about $\mu$G from the core ($r_{1000}$) of the clusters (above $5\times 10^{14}M_{\odot}$) as expected from many Faraday rotation observations of galaxy clusters \citep{Eilek_2002ApJ,Govoni_2004IJMPD}. Whereas, groups are found to have 10s of nano-Gauss to sub $\mu$G magnetic field with a considerable amount of fluctuations in their values (see Fig.~\ref{mag-field}, Panel 2). 

\subsection{Modelled radio emissions}\label{Sim-obs-comp}

\begin{figure}
%\hspace{-1cm}

\includegraphics[width=0.52\textwidth]{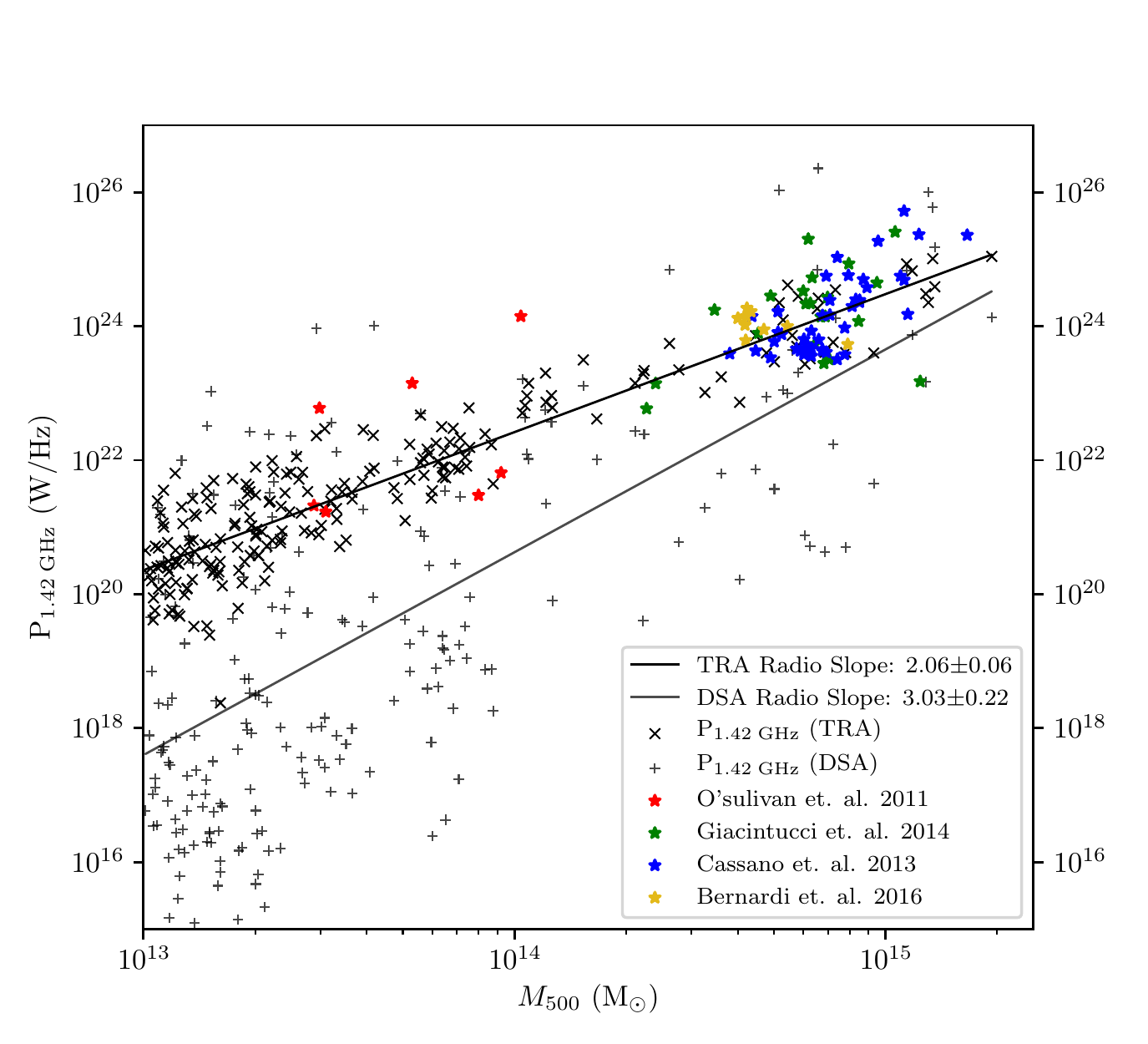}\hspace{-.5cm}
\includegraphics[width=0.52\textwidth]{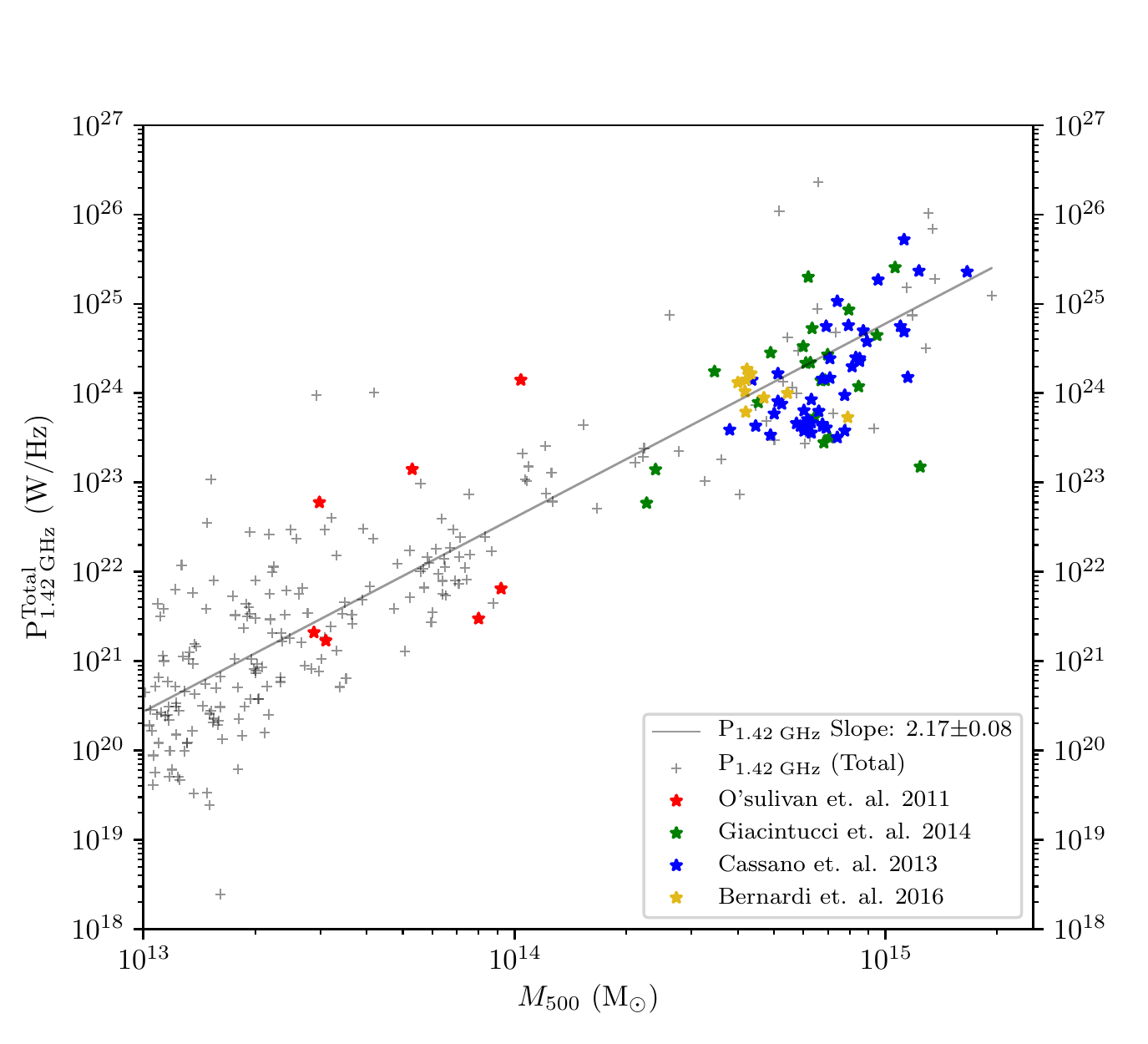}

\caption{{\bf Panel 1} Over-plotted observation data on modelled radio power from DSA \& TRA models from simulated groups and clusters. {\bf Panel 2} Same set of observed data has been plotted over the computed total radio power. }\label{ovelplots}
\end{figure}

Our radio emission models described in section~\ref{DSA-compt} and section~\ref{TRA-compt} have been implemented to compute 1.4 GHz radio synchrotron emission power from both DSA and TRA electrons in the chosen simulated objects. From an observational point of view, we have also performed a band integration over the central frequency 1.4 GHz with the band size of 32 MHz (e.g. GMRT L band) to compute radio synchrotron power from both DSA and TRA mechanisms. The magnetic fields computed using the formulation described in section~\ref{mag-field-compt} are used in these calculations. To verify and constraint our model from observations, we need to understand the bulk properties of the sources for which adequate observations are available. Since, radio halos have been studied adequately, at least at large masses, we have computed mainly the halo emissions for the comparison. Radio halo emission in the clusters are usually found to be of sizes about a Mpc \citep{Feretti_2008LNP,Giovannini_2009A&A,Paul_2014ASInC} in the core region i.e. their linear extension can be well approximated to their $r_{1000}$ radius. So, we compute the radio emissions from this region only. As we have seen in the section~\ref{mach-gr-cl} that shocks are also present in plenty inside the core, we understood that while computing the radio halo emission from cluster cores, the emission due to DSA cannot be neglected. 

Results show that the radio power of the sources vary from $10^{15}\;\rm{Watt\;Hz^{-1}}$ to $10^{26}\;\rm{Watt\;Hz^{-1}}$ with different mass scaling for DSA and TRA. We have over-plotted observed data of radio power at 1.4 GHz from \cite{Bernardi_2016MNRAS,Giacintucci_2014ApJ,Cassano_2013ApJ,Cassano_2007MNRAS,O'Sullivan_2011ApJ} on our modelled radio power plot. It has been noticed that for many objects, radio emissions due to DSA electrons are equally important as the TRA for emissions from the central halo (within $R_{1000}$), especially from the objects with mass greater than $10^{14} M_{\odot}$. While, in lower mass systems, TRA turns out to be more effective (see Fig.~\ref{ovelplots}, Panel 1). From the observations, we have very few points below mass of $5\times 10^{14} M_{\odot}$ and the plotted few red and green points were thought to be outliers \citep{Giovannini_2011A&A}, considering the scaling laws of clusters beyond $5\times 10^{14} M_{\odot}$ as the reference which follow a much steeper mass scaling of about 4 \citep{Cassano_2013ApJ,Yuan_2015ApJ}. But, when we add up the radio powers from DSA and TRA and over-plotted again the observed data, the thought outliers perfectly fell along the same line indicating a possible misinterpretation by an incomplete and subset of observed data (above $5\times 10^{14} M_{\odot}$) in the literature and thus validates our theoretical model and allowed us to extend it to the lower mass objects as well. Here we report a new predicted correlation of $M_{500} \propto P_{1.4\;GHz}^{2.17\pm0.08}$ with total radio power (DSA+TRA) at 1.4 GHz which extends from mass $10^{13} M_{\odot}$ to more than $10^{15} M_{\odot}$. From the Figure~\ref{ovelplots}, Panel 2, it is evident that our targeted objects i.e. groups are having much higher radio power than expected from observed cluster scaling and emission from groups is mainly dominated by TRA, making them detectable with the advanced radio telescopes.

\subsection{Correlation among modelled radio and X-ray emissions}\label{x-ray-comp}

\begin{figure}
%\hspace{-1cm}
\includegraphics[width=0.52\textwidth]{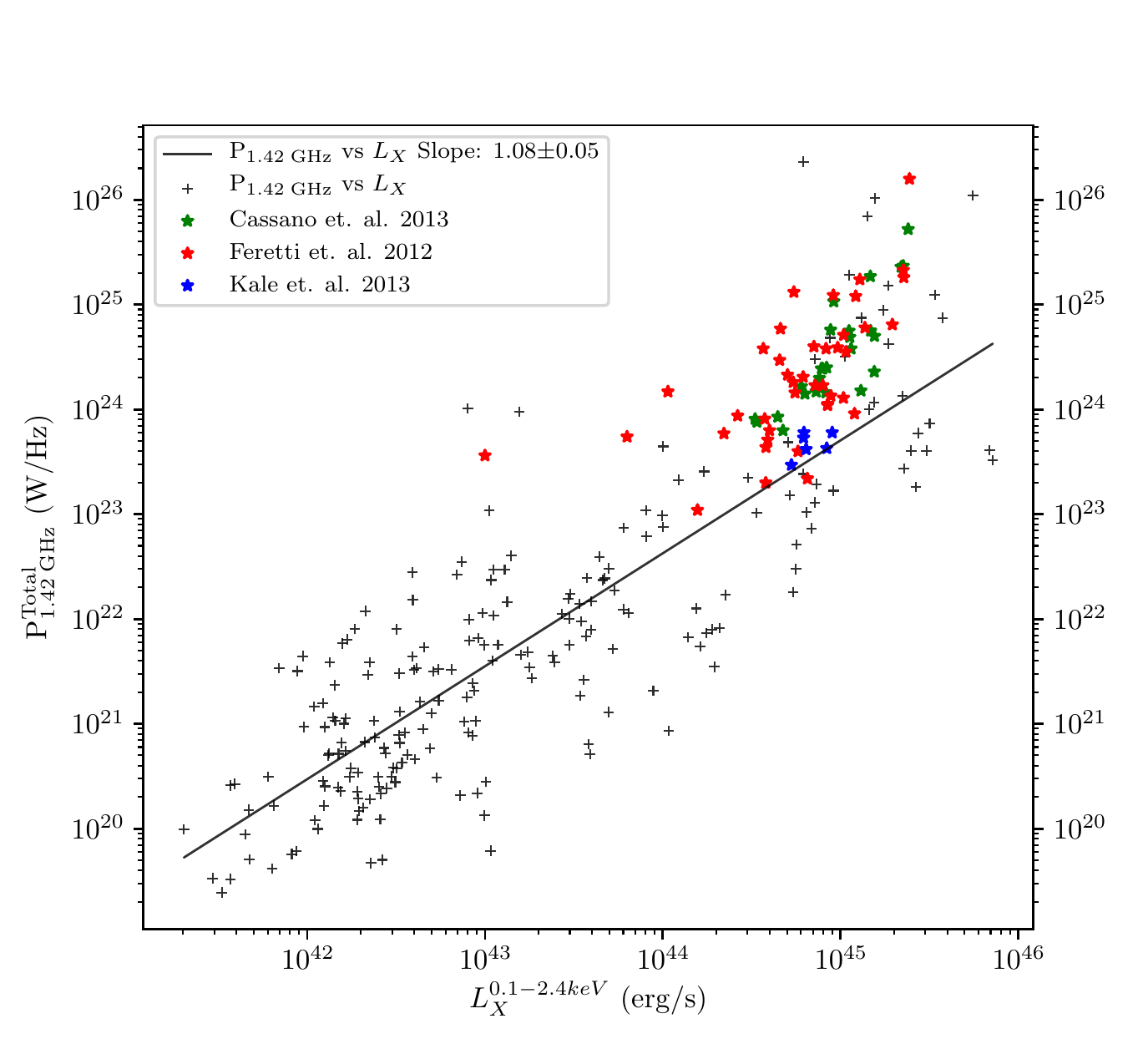}
\caption{Over-plotted observation data on modelled X-ray luminosity vs radio power (DSA \& TRA) from simulated groups and clusters.}\label{ovelplots-lx-lrad}
\end{figure}

Shocks emerged out of mergers are responsible for heating the medium as well as acceleration of particles, increment of the turbulence, and amplification of magnetic fields in the cluster medium (e.g., \citet{Carilli_2002ARA&A,Dolag_2002A&A,Bruggen_2005ApJ, Subramanian_2006MNRAS,Ryu_2012SSRv,Paul_2011ApJ}). Even though, X-ray emission takes place due to thermal bremsstrahlung of hot plasma (1-10 keV) in the  IGM and radio emissions takes place due to synchrotron emission from gyrating charged particles accelerated by shocks and turbulence in the shock and turbulence amplified magnetic fields, a strong correlation has been observed to sexists among them \citep{Colafrancesco_2014A&A,Cassano_2013ApJ}. Since the mechanisms are different (one is thermal and other in non-thermal), the only possible explanation for this correlation can be attributed to the mergers of clusters. Since, the same merging events thermalize as well as accelerate particles that eventually emits X-ray (thermal) and radio (non-thermal), total luminosity possibly be correlated. But, the correlation of radio power at 1.4 GHz ($P_{1.4\;GHz}$) and Xray ($L_X$, 0.1-2.4 keV) reported so far is for a very limited sample set and are only for large mass objects ($>$ 10$^{14}M_{\odot}$), could not reveal anything about low mass `galaxy groups'. In Figure~\ref{ovelplots-lx-lrad}, we have plotted our modelled radio emissions (see section~\ref{radio-model}) against the modelled X-ray emission (computed using {\sc cloudy} code \citep{Ferland_1998PASP}, (for detailed parameters see \citet{John_2018arXiv})) from our simulated sample in the mass range 10$^{13}$ -10$^{15}M_{\odot}$. The observed data from various available literature have been over-plotted in the same plot for confirming the accuracy of our model. It shows, similar to radio power mass scaling, the points that were thought to be outliers by 
\citet{Giovannini_2011A&A,Giovannini_2009A&A} in their $L_X - P_{1.4\;GHz}$ correlation plot, came well under our computed correlation line as shown in Figure~\ref{ovelplots-lx-lrad}. This result further validates of our model. A new correlation scale of $L_X \propto P_{1.4\;GHz}^{1.08\pm0.05}$ has been found in our study. Since, our model fits well to all the available observed data (see Fig.~\ref{ovelplots-lx-lrad}), it has enabled us to extend the computed correlation towards the lower mass which so far has remained unexplored.

\subsection{Computing radio emissions from selected SDSS groups}\label{SDSS-radio-model}

Our simulated models (section~\ref{mag-field-compt},~\ref{DSA-compt}~and~\ref{TRA-compt}) compared and constraint with various available observations (as discussed in section~\ref{mag-field-res},~\ref{Sim-obs-comp}~and~\ref{x-ray-comp}) have been implemented to real set of objects derived from Sloan Digital Sky Survey (SDSS) data (details of the data selection in section~\ref{obs-data}) .
The most important factor for computing magnetic field and radio emission is turbulence in the medium. In our dataset of SDSS, we have data for velocity dispersion that are computed from the line-of-sight velocities of all detected member galaxies inside individual groups \citep{Tempel_2014A&A}. Velocity dispersion is an indication of turbulence in the medium as has been discussed already (section~\ref{turb-in-gg-gc}). Further, SDSS dataset \citep{Tempel_2014A&A} contains mass, velocity dispersion etc. required parameters only at over-density 200 (i.e. $r_{200}$), but mostly mass and X-ray emissions are reported up to over-density 500 (i.e. $r_{500}$) in most of the literature. Also, radio halo emissions that usually observed is contained within the core of the structure i.e. about over-density 1000 (i.e. $r_{1000}$), accordingly, we have scaled required parameters using our simulations data. We have finally computed radio emissions implementing the models described in section~\ref{DSA-compt}~and~\ref{TRA-compt}. 

\begin{figure}

\includegraphics[width=0.52\textwidth]{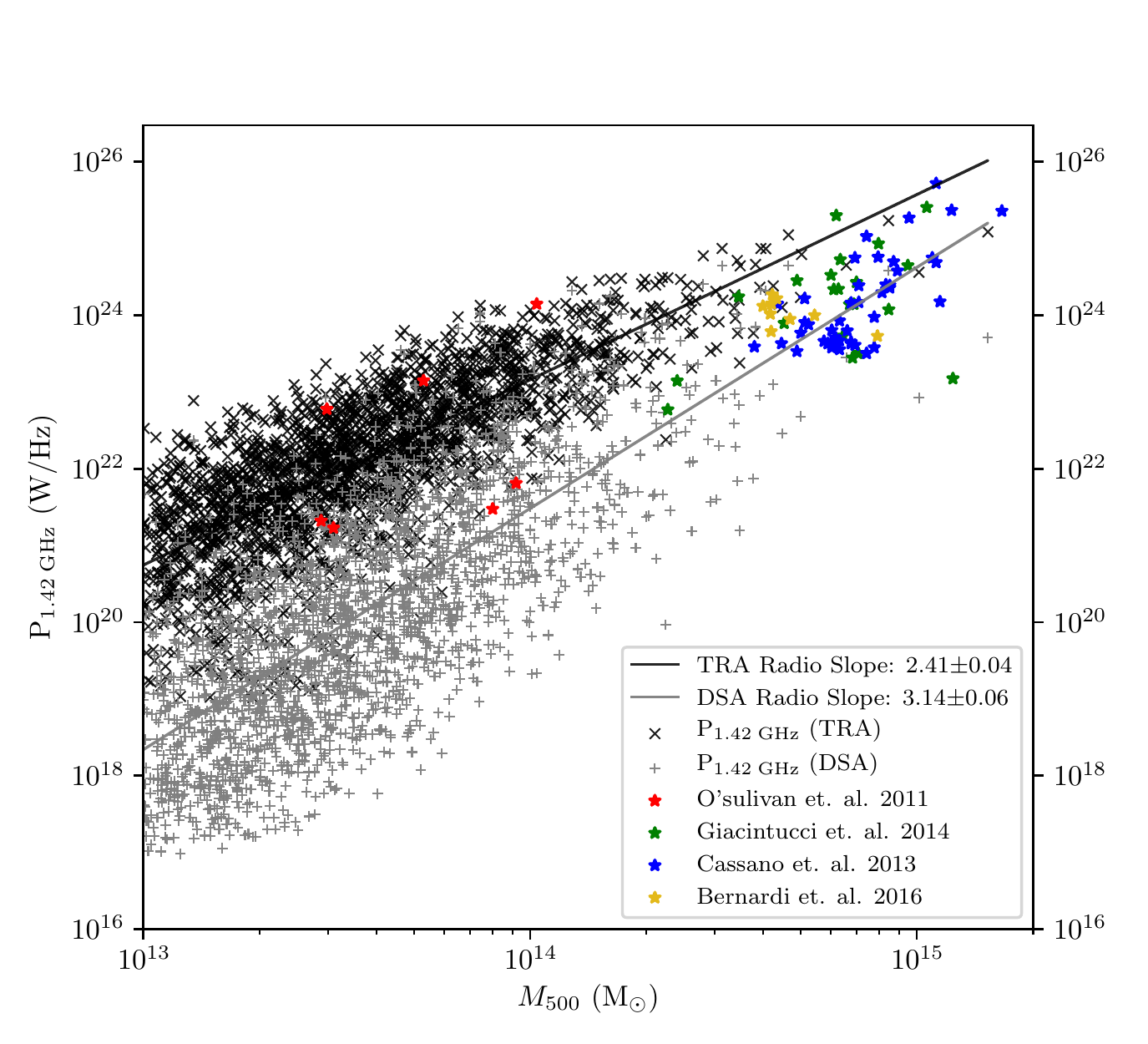}

\caption{Modelled radio power (DSA \& TRA) from SDSS groups and clusters.}\label{radio-pow-sdss}

\end{figure}

Estimated radio emission power from the selected samples of SDSS galaxy groups and observations are over-plotted similar to Figure~\ref{ovelplots}. It has been found that our computed values for SDSS data also fairly match with the observations (see Fig.~\ref{radio-pow-sdss}). The only difference is the slope of TRA against mass, which is little steeper than the simulated one (see Fig.~\ref{ovelplots}). For computation of expected radio emissions from SDSS groups, we had to use the averaged values which miss the possible fluctuations inside the volume explaining the slight difference in slope observed compared to the simulated data.

\subsection{Prospect of detection of SDSS galaxy groups }

\begin{figure}

\includegraphics[width=0.52\textwidth]{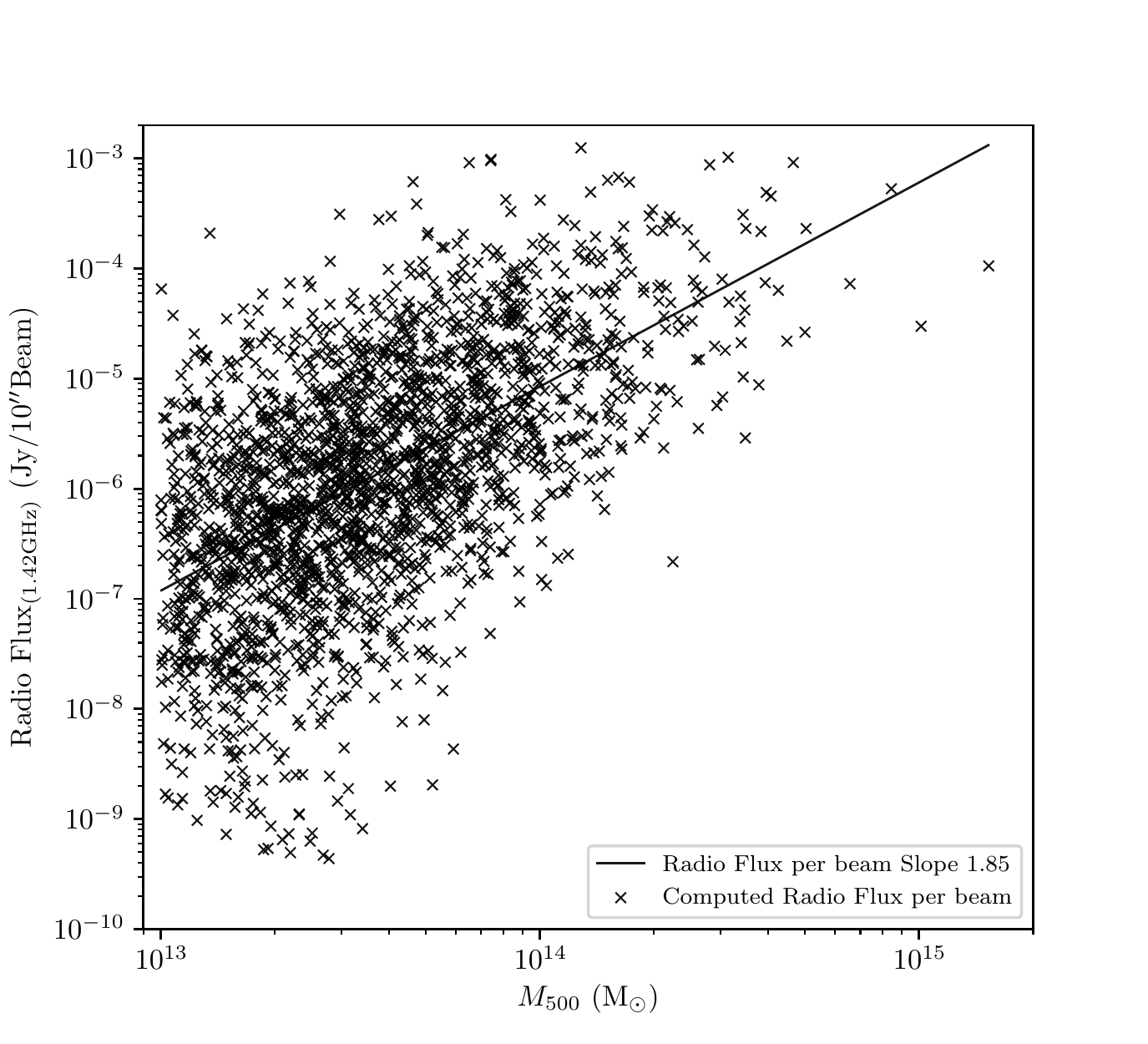}
  \caption{Total radio flux (DSA+TRA) per beam (10{\arcsec}) from our selected SDSS data set plotted vs the object mass ($M_{500}$).} \label{radioFlux}
  \vspace{-0.2cm}
\end{figure}

Applying our theoretical model of radio emission to the selected SDSS groups, we have estimated the 1.4 GHz flux per beam taking a beam size of (10{\arcsec}). The computed flux varies in the range of sub $\mu$Jy to few 10s of $\mu$Jy per beam (see Fig.~\ref{radioFlux}). The radio emission produced, travel to reach the earth surface. Since, radiation goes spherically out, flux from the source goes down by a factor of $L_d^2$ where, $L_d$ is the luminosity distance of each of these objects. Further, we have considered the angular extension of the radio emitting part i.e. $r_{1000}$ of each object to compute the flux per beam. Expected 1.4 GHz flux on earth for these groups are computed in Jy/beam where 1~Jy = $10^{-26}\; \rm{Watt\;m^{-2}}$ (beam size is 10$^{\arcsec}$) and plotted in the Figure~\ref{radioFlux}. Many of the low mass objects show a promising level of fluxes that are even within the range of few existing or upgraded radio telescopes. We have further created data set of objects with expected radio flux above 100$\mu$Jy/beam (see Table will be available with the published version of the paper), in view of possible interest in astronomy community for observing them. With a 100$\mu$Jy/beam limit, we can expect to observe 75 objects i.e. about 3\% of 2300 selected SDSS groups in our sample. By pushing the detection limit by 10 times i.e. to 10 $\mu$Jy, we can increase our detection success rate to 21\%  (about 490 objects) which seems very promising with the view of upcoming telescopes such as SKA. In this context, we should mention that one of the very small mass (about $3\times 10^{13} M_{\odot}$) object has already been detected by us at GMRT 610 MHz and will be reported in a separate article (Paul S., et al., in preparation). Our predicted value of mean radio flux is 70$\mu$Jy/beam, when observed value is about 100$\mu$Jy/beam. And further detection of smaller groups (about $10^{13} M_{\odot}$) having temperature less than 1 keV will reveal the WHIMs and the missing baryons along with the nano-Gauss magnetic fields. 

\subsection{Limitations of this study}

Our computations have been performed on a hydrodynamic set-up but, many of the parameters required for computing magnetic field and radio synchrotron emissions are purely of magnetohydrodynamic (MHD) in nature. So, for more realistic results we need to perform MHD simulations. Further, particle acceleration models are used on a set of time frozen simulation outputs, no radiation transport has been considered, and synchrotron spectral ageing has also been ignored, thus time evolution of the parameters are missing in our calculations. While implementing our simulated models to observed SDSS objects, we had to use scaling relations to compute the velocity dispersion and density. Since, SDSS parameters are available as averaged over the total volume of the objects, our computed values miss the spatial variation information. All these may lead to some amount of over or underestimation of the computed values.

\section{Summary and conclusion}\label{sum}

In this study, for the first time, a comprehensive radio emission model for LSS has been presented. Our model includes synchrotron radio emission from IGM electrons accelerated by both DSA and TRA. Computed radio power in our study matched well with the available data of radio emissions at a wide range of mass and also verified with the $P_{1.4\;GHz}-L_X$ correlation plot. This is worth mentioning that our models (see section~\ref{radio-model}) are very robust and we hardly need to tweak or constrain any parameters to fit it with the observations. This enabled us to correctly extrapolate to estimate possible radio emissions from low mass objects. We report a detectable amount of radio emissions from low mass objects with the high sensitive available and upcoming radio telescopes such as uGMRT, SKA, ALMA and so on. So, the major findings from this study are as follows.

\begin{itemize}
\item We have modelled saturated magnetic field in groups and clusters taking very simple assumptions as described in section~\ref{mag-field-compt}. Our model has been able to well reproduce the observed radial profile of coma cluster (see Fig.~\ref{mag-field}). We report a 10s of nano-Gauss to sub $\mu G$ magnetic field in groups and any detection of such groups will have a significant role in constraining turbulent dynamo model in explaining the origin of cosmic magnetism. 

\item We have noticed that the radio halo emission observed so far in the galaxy clusters cannot be explained only with TRA model as it is usually done. We found a significant role of merger shocks even inside the $r_{1000}$ of the clusters and we report that a combined radio power of DSA and TRA can only fit well to all the observed objects. It has also been noticed that the galaxy groups are clearly dominated by TRA radio emissions and do not fall in the current mass-radio power scaling law derived from an incomplete data set as given in \citet{Cassano_2013ApJ,Yuan_2015ApJ} rather, follow a flatter slope of $\alpha={2.17\pm0.08}$ as found in our study. Similar observation can be made about $L_X - P_{1.4\;GHz}$ correlation where we found a much flatter slope of $\beta=1.08\pm0.05$. 

\item The radio emission model devised using simulation (see section~\ref{radio-model}) has been implemented to the real objects obtained from SDSS galaxy group catalogue in our study. We found many smaller groups have the radio flux much above the detection limit of currently available telescopes. So, with a proper strategy, a considerable number of (about 3 \% of the selected groups from SDSS list) of low mass objects should show up in radio soon. Indeed, we have detected radio halo emissions from one of the very small mass (about $3\times 10^{13} M_{\odot}$) SDSS object (will be reported in a separate article in Paul S., et al., in preparation). Further, with the upcoming telescopes such as SKA, we will be able to detect about 20\% of the total selected SDSS groups. This also raises the hope of detection of WHIMs and can shed light on the outstanding missing baryon problem.
\end{itemize}

\section*{Acknowledgements}
This project is funded by DST-INSPIRE Faculty Scheme (IFA-12/PH-44), Govt. of India. SP likes to thank Prof. Marcus Bru\"ggen of Hamburger Sternwarte, Hamburg University for his valuable suggestions and discussions and to DESY, Hamburg for funding a visit to Hamburger Sternwarte as SFB Fellow to initiate collaborative research projects. We are thankful to The Inter-University Centre for Astronomy and Astrophysics (IUCAA) for providing the HPC facility. Computations described in this work were performed using the Enzo code developed by the Laboratory for Computational Astrophysics at the University of California in San Diego (http://lca.ucsd.edu), and data analysis is done with the yt-tools (http://yt-project.org/ \citep{Turk_2011ApJS}).

\bibliographystyle{apj}
\bibliography{Radio-groups-apj.bib}

%% This command is needed to show the entire author+affilation list when
%% the collaboration and author truncation commands are used.  It has to
%% go at the end of the manuscript.
%\allauthors

%% Include this line if you are using the \added, \replaced, \deleted
%% commands to see a summary list of all changes at the end of the article.
%\listofchanges

\appendix

\section{Resolution study}

With the cosmological and simulation parameters described in section~\ref{simu}, we have simulated realisations with 6 levels of total (uni-grid + AMR) refinement leading to a resolution of about 30 kpc i.e. the RefRES. For the resolution study, apart from this RefRES, we have simulated two other sets of data with a lower resolution (`LowRES' hereafter) reaching $\sim$ 60 kpc with total 5 AMR and a higher resolution with total 7 AMR (`HighRES' hereafter) $\sim$ 15 kpc by keeping the other parameters same. We have also produced two simulations having two different root Grid resolutions besides the `RefRES'. High resolution root grid simulation is done with 128$^3$ and is named as `RootHighRES' and low resolution one is with 32$^3$ root grids and called as `RootLowRES'. But, in these last two simulations, the final resolution is same as the RefRES i.e. about 30 kpc at the highest resolution level.

\begin{figure}
\includegraphics[width=0.5\textwidth]{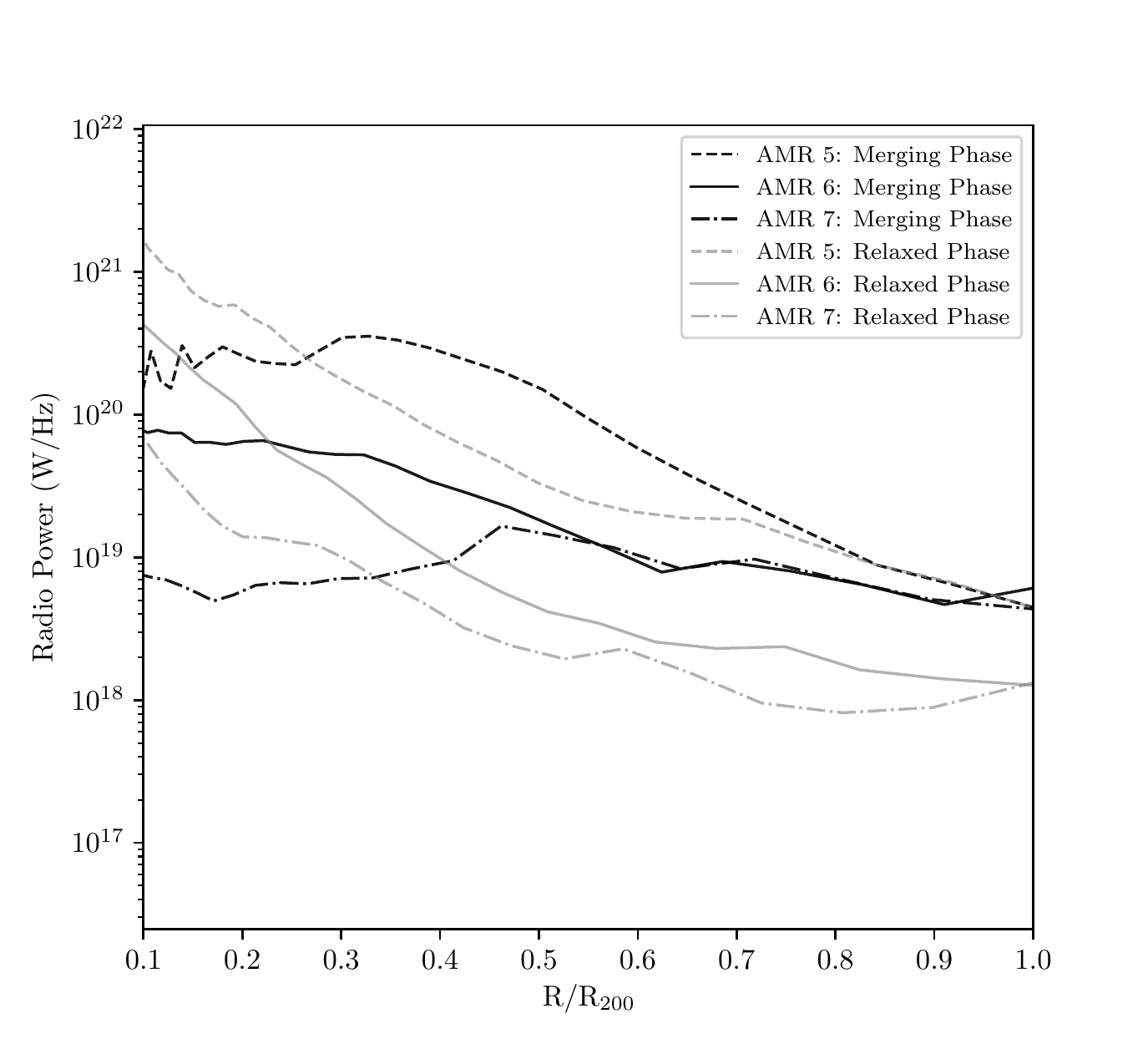}
\includegraphics[width=0.5\textwidth]{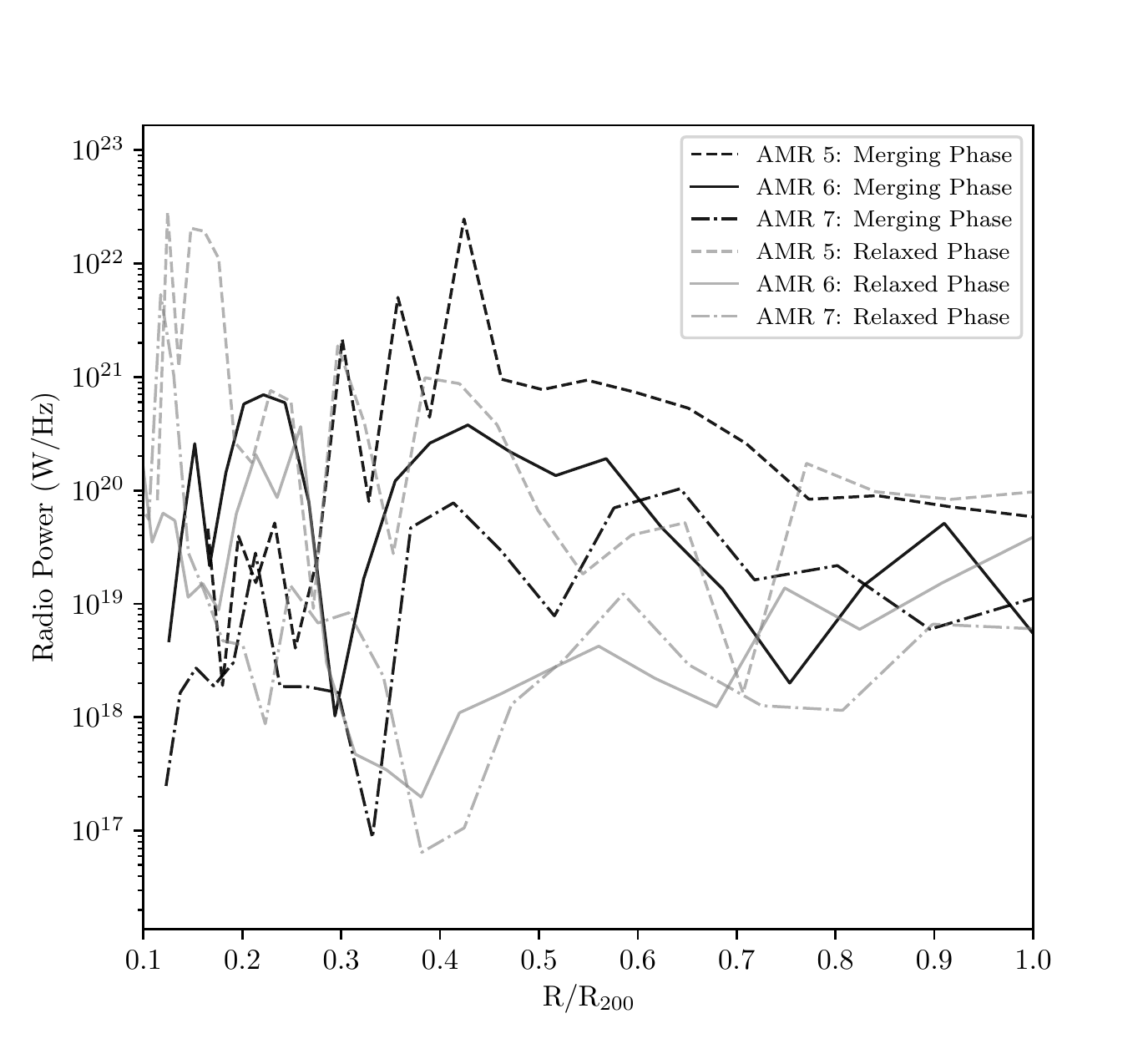}
\caption{Computed radio power (Panel 1 is TRA and Panel 2 is DSA) have been plotted against normalised radius (normalised to $r_{200}$) for merging and relaxed states, of a galaxy cluster with final mass about 10$^{15} M_{\odot}$ and for three resolutions namely LowRES, RefRES and HighRES (as indicated in the legend respectively)}\label{AMR-res}
\end{figure}

In Figure~\ref{AMR-res}, radial variations of radio power due to TRA (Panel 1) and DSA (Panel 2) respectively have been plotted for merging (Black) and relaxed phase (Grey) of the cluster $CL_2$. It can be noticed that RefRES simulation is almost same as the HighRES resolution with some deviation in the central region of the cluster, though, LowRES data are far away. Radio luminosity varies smoothly in case of TRA electrons as this comes from a bulk property of the cluster whereas fluctuations are more in case of DSA as it has resulted from shocks which is a transient effect which is strongly dependent on resolution.

\begin{figure}
\includegraphics[width=0.5\textwidth]{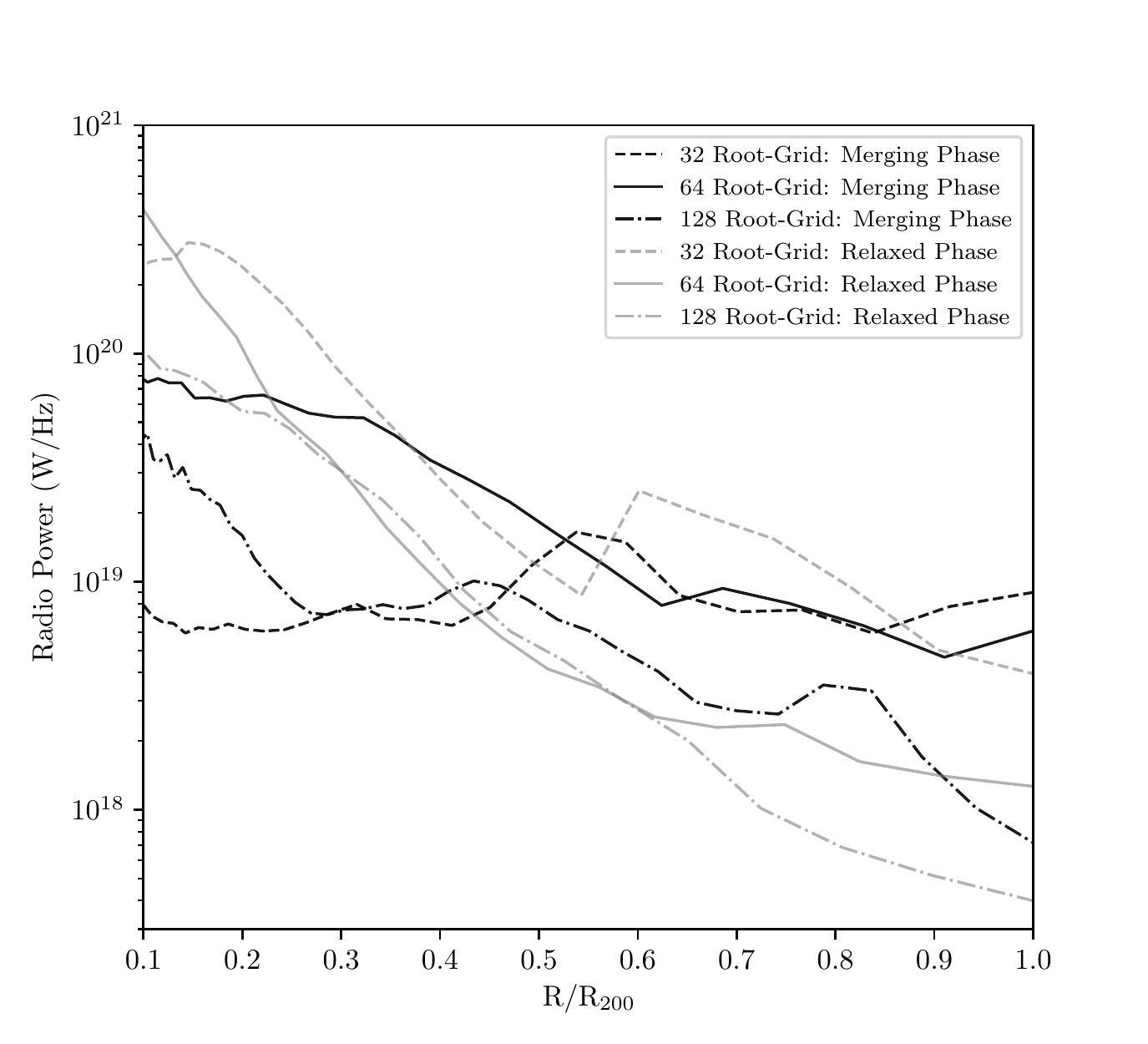}
\includegraphics[width=0.5\textwidth]{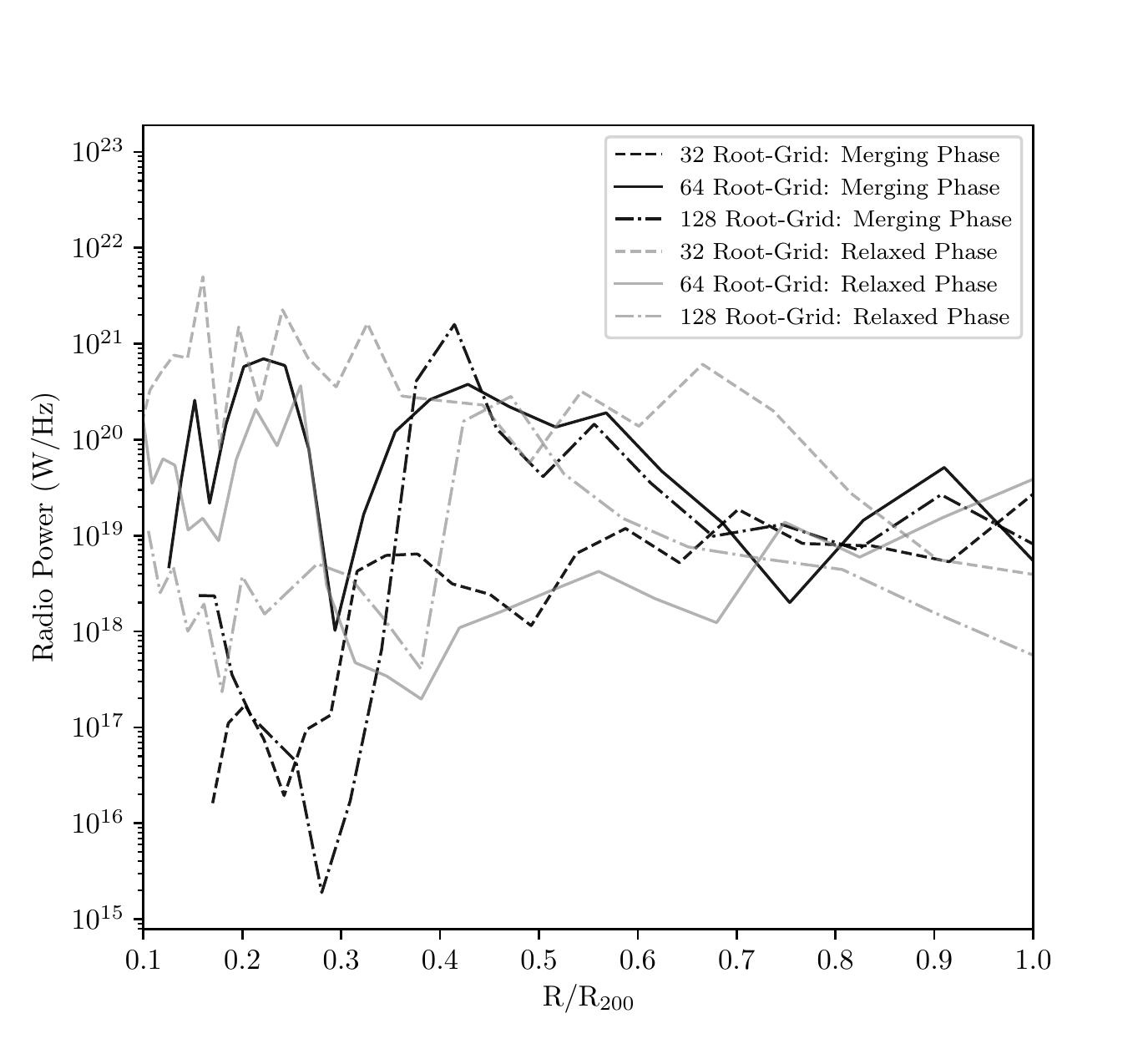}
\caption{Plot of radial (normalised radius, as in Fig.~\ref{AMR-res}) radio power (Panel 1 is TRA and Panel 2 is DSA) for the same dynamical phases for different root grid resolution simulations with the same final resolution of about 30 kpc.}\label{root-res}
\end{figure}

Another two sets of data with different root grid resolution namely RootLowRES and RootHighRES have been simulated. We have further chosen appropriate merging and non-merging states and plotted the radio emissions again from both TRA and DSA electrons. Though we observe a reasonable convergence during non-merging phase, merging phases show a greater deviation at least in some part of the objects where shock has propagated (see Fig.~\ref{root-res}, Panel 1). Again, the discrepancy can be attributed to the resolution sensitivity of dynamical effects but the total emissions from the whole object seems not changing much. This indicates that though there could be a little time shift in the parameters but has no gross deviation from computed magnitude of the parameters when measured by compensating the time. 

These results thus show that our simulated parameters are almost converging with the resolution that we took as the reference set of simulations i.e. about 30 kpc with 6 levels of refinement. For further details of resolution study of our data sets, we suggest to go through \citet{Paul_2017MNRAS,John_2018arXiv}.

\section{Table of objects with predicted radio flux of 50 $\mu$Jy/beam}\label{Grp-table}

%\begin{table*}
%\tablewidth{1.0\columnwidth}

%\captionsetup{width=14cm}
%\onecolumn

\end{document}